\documentclass{aa}
\usepackage{longtable}
\usepackage{graphicx}
\usepackage{txfonts}
\usepackage{subcaption}
\usepackage{placeins}
\usepackage{amsmath}
\usepackage{booktabs}
\usepackage{multirow}
\usepackage[colorlinks=true,citecolor=blue,linkcolor=blue,urlcolor=blue]{hyperref}
\usepackage{capt-of}
\newcommand{\HII}{H\,{\sc ii}}
\begin{document}
\titlerunning{Gas-Phase Metallicity Gradients in MaNGA Low-Mass Galaxies}
\authorrunning{Dasgupta}

\title{Internal Structure, Not Environment, Shapes Metallicity Gradients in Low-Mass Galaxies from MaNGA}

\author{
Anuroop Dasgupta\inst{1,2,3}\corrauth{anuroop.dasgupta@mail.udp.cl}
\and
Namrata Roy\inst{4}
\and
Saumyadip Samui\inst{5}
\and
Ritaban Chatterjee\inst{5}
\and
Fuyan Bian\inst{1}
}
\institute{
European Southern Observatory, Alonso de C\'ordova 3107, 
Casilla 19001, Vitacura, Santiago 19, Chile
\and
Instituto de Estudios Astrof\'isicos, Facultad de Ingenier\'ia 
y Ciencias, Universidad Diego Portales, Av. Ej\'ercito 
Libertador 441, Santiago, Chile
\and
Department of Physics, Presidency University, 86/1 College 
Street, Kolkata, 700073, India
\and
School of Earth and Space Exploration, Arizona State 
University, Tempe, AZ 85287, USA
\and
School of Astrophysics, Presidency University, 86/1 College 
Street, Kolkata, 700073, India
}
\date{Received July 1, 2026; Revised August 31, 2026; Accepted September 29, 2026}

\abstract
% Context
{Dwarf and low-mass galaxies are highly 
sensitive to stellar feedback and gas flows, 
yet their spatially resolved gas-phase 
metallicity remains poorly characterised 
for lack of large homogeneous integral field 
spectroscopy samples.}
% Aims
{We measure radial gas-phase metallicity 
gradients for the largest MaNGA low-mass 
sample to date and investigate the physical 
processes governing their chemical 
enrichment.}
% Methods
{We assemble 382 star-forming low-mass 
galaxies from MaNGA Data Release 17, 
spanning $7.2 \leq \log(M_{\star}/M_{\odot}) 
\leq 9.1$ and $z \leq 0.05$. Spectra are 
stacked in elliptical radial bins and 
gas-phase metallicity is derived from the 
N2 index. After removing the primary 
dependence on local stellar mass surface 
density $\Sigma_{\star}$, we define a 
residual metallicity $\Delta Z$ to isolate 
secondary dependencies.}
% Results
{Metallicity gradients are flat on average 
(median slope $-0.007$\,dex\,$R_{\rm 
e}^{-1}$). The residual gradients correlate 
most strongly with light concentration 
(Pearson $r = 0.304$, $r^2 = 0.092$, 
$p < 0.001$) and gas velocity dispersion 
($r = 0.204$, $p < 0.001$), though both 
are modest predictors accounting for at 
most 9 per cent of the variance. The 
velocity dispersion result should be 
interpreted with caution given its 
sensitivity to spatial resolution. Stellar 
mass shows no significant Pearson 
correlation, large-scale environment shows 
no significant correlation in either test, 
and star formation rate shows only a 
marginal correlation ($r = 0.106$, 
$p = 0.042$).}
% Conclusions
{The concentration correlation is robust 
to measurement uncertainties and persists 
when a stricter spatial resolution cut is 
applied, suggesting that internal 
structural properties may play a role in 
shaping the radial distribution of metals 
in low-mass galaxies. However the modest 
effect sizes indicate that no single 
parameter can fully account for the 
observed diversity in metallicity 
gradients, and larger samples will be 
needed to fully characterise the drivers 
of chemical enrichment in this mass 
regime.}
{}
\maketitle
\nolinenumbers
\keywords{galaxies: dwarf -- 
galaxies: abundances -- 
galaxies: evolution -- 
galaxies: ISM -- 
techniques: spectroscopic -- 
surveys}

\section{Introduction}
\label{sec:intro}

The chemical evolution of galaxies is one of the central 
problems in extragalactic astrophysics. As galaxies form 
stars from their gas reservoirs, successive generations 
of stellar nucleosynthesis progressively enrich the 
interstellar medium with heavy elements. At the same time, stellar 
feedback in the form of supernova driven winds can expel 
enriched material into the circumgalactic medium, and 
accretion of pristine gas from the cosmic web can dilute 
the existing metal content \citep{Maiolino2019}. The 
net result of these competing processes is encoded in 
the gas-phase metallicity of galaxies, commonly measured 
as the oxygen abundance $12 + \log(\mathrm{O/H})$, which 
 serves as a powerful integrated tracer of a 
galaxy's star formation history, gas inflow and outflow 
history, and chemical enrichment \citep{Maiolino2019}. 
The most direct manifestation of this enrichment at the 
population level is the mass-metallicity relation 
\citep[MZR;][]{Tremonti2004}, which establishes that 
more massive galaxies are on average more metal rich, 
a trend that has now been confirmed across a wide range 
of redshifts and environments \citep{Mannucci2010, 
KewleyEllison2008}. While the MZR captures the global 
enrichment level of galaxies, it cannot reveal how 
metals are distributed spatially within them. Spatially 
resolved measurements of gas-phase metallicity, made 
possible by integral field spectroscopy, have opened 
a new window into the physical processes that drive 
chemical enrichment across the disks of galaxies 
\citep{Sanchez2020}.

The radial gradient of gas-phase metallicity within a galaxy, 
typically quantified as the slope of the 
$12 + \log(\mathrm{O/H})$ profile, encodes information about 
the dominant physical processes shaping chemical evolution. Negative 
gradients, where the central regions are more metal rich 
than the outskirts, are the natural expectation from 
inside-out disk growth, in which star formation proceeds 
first in the central regions and gradually extends 
outward over time \citep{Matteucci1989, Boissier1999}. 
The first systematic measurements of metallicity 
gradients in nearby spiral galaxies were obtained from 
observations of \ion{H}{ii} regions, establishing that 
negative gradients of typically $-0.1$ to 
$-0.2$\,dex\,$R_{\rm e}^{-1}$ are common in massive 
late-type galaxies \citep{Zaritsky1994, Moustakas2010}. 
The advent of integral field spectroscopy surveys brought 
a dramatic improvement in both sample size and spatial 
coverage. Using data from the CALIFA survey, 
\citet{Sanchez2014} showed that the metallicity gradient 
of isolated spiral galaxies is remarkably uniform at 
approximately $-0.1$\,dex\,$R_{\rm e}^{-1}$, independent 
of stellar mass, morphology, and absolute magnitude. 
Subsequent work with MaNGA confirmed 
this picture for massive star-forming 
galaxies 
($\log(M_\star/M_\odot) \gtrsim 10$) 
while revealing that gradients flatten 
toward both lower masses 
($\log(M_\star/M_\odot) \lesssim 9$) 
and higher masses 
($\log(M_\star/M_\odot) \gtrsim 11$) 
\citep{Belfiore2017}, and that beam smearing from the 
point spread function can artificially flatten gradients 
in poorly resolved galaxies \citep{Acharyya2020}. These 
surveys also established a spatially resolved 
mass-metallicity relation, in which local gas-phase 
metallicity correlates with local stellar mass surface 
density $\Sigma_{\star}$, indicating that chemical 
enrichment is regulated on local scales within galaxies 
\citep{RosalesOrtega2012, Sanchez2013, BarreraBallesteros2016}. 
Metallicity gradients are also sensitive to dynamical 
perturbations. Merging and interacting systems bring in 
fresh, metal-poor gas that can drive the metallicity 
gradient to be more negative, while the radial mixing of 
metals over time and feedback-driven outflows tend to 
redistribute metals across the interstellar medium, making 
the gradient flatter or even slightly positive 
\citep{Rupke2010, Kewley2010}. The balance of these 
competing processes sets the observed gradient, motivating 
the search for the physical drivers of gradient diversity 
that is the focus of this work.

While metallicity gradients in massive galaxies are 
relatively well characterised, the situation is 
considerably less clear for dwarf galaxies, i.e., with stellar masses below 
$10^{9}\,M_{\odot}$. Dwarf galaxies occupy a 
qualitatively different physical regime from their 
massive counterparts. Their shallow gravitational potential wells make them highly 
susceptible to supernova driven feedback, which can launch 
large scale galactic winds capable of expelling enriched gas 
from the entire disk rather than merely redistributing it 
radially \citep{Veilleux2005, Samui2008, Tolstoy2009}. As a consequence, 
the inside-out growth scenario that 
naturally produces negative metallicity gradients in 
massive galaxies may be fundamentally disrupted in 
dwarfs, where feedback driven outflows and inflows can 
mix and even invert intrinsic metallicity gradients 
built up by spatially concentrated star formation. 
Cosmological hydrodynamical simulations of dwarf 
galaxies generally predict weak gas-phase metallicity 
gradients, with large galaxy-to-galaxy scatter driven 
by the bursty and irregular nature of star formation 
in shallow potential wells \citep{Mercado2021}. Furthermore, dwarf galaxies 
are more susceptible to environmental processes such 
as tidal interactions and ram pressure stripping, 
which can further redistribute gas and metals \citep{Boselli2022}.

Observational studies of spatially resolved 
gas-phase metallicity in dwarf galaxies have 
until recently been limited to small samples. 
The advent of integral field spectrographs 
on large telescopes enabled the first 
resolved metallicity maps of dwarf galaxies, 
though initial studies were necessarily 
limited to small numbers of objects. 
\citet{Grossi2020} used integral field 
spectroscopy of two Virgo cluster dwarf 
galaxies ($\log(M_{\star}/M_{\odot}) \sim 
8$--$9$) to derive metallicity maps and 
radial profiles, finding inverted, positive 
gradients that they attributed to recent 
merging and the inflow of metal-poor gas. 
At the low mass end, \citet{Belfiore2017} 
found that the metallicity gradients in 
MaNGA galaxies lie roughly flat at 
$\log(M_{\star}/M_{\odot}) \sim 9.0$, 
which suggests that the flattening continues 
toward lower masses into the dwarf regime. 
Recent dedicated studies using MUSE and 
SAMI have begun to extend this to lower 
masses \citep{Poetrodjojo2021, Li2025, 
Vale2025}, but samples remain small or 
limited to $\log(M_{\star}/M_{\odot}) 
\gtrsim 8.5$. Despite these advances, the 
physical drivers of metallicity gradient 
diversity in low-mass galaxies remain poorly 
constrained, in particular the relative 
roles of internal feedback mechanisms and 
external environmental processes. These can 
be probed observationally through proxies 
such as light concentration, morphology, 
gas velocity dispersion, star formation 
activity, and the large scale environment. 
A large, homogeneous sample of low-mass 
galaxies with reliable spatially resolved 
metallicity measurements is needed to make 
progress on these questions.

In this paper we present the largest sample of dwarf 
galaxies assembled from the complete Data Release 17 of the 
MaNGA integral field spectroscopy survey. We select 382 
star-forming dwarf galaxies with stellar masses 
$M_{\star} < 10^{9.1}\,M_{\odot}$, absolute magnitudes 
$M_{g} > -18.5$, and redshifts $0.01 \leq z \leq 0.05$, 
closely following the selection criteria of the MaNGA Dwarf 
Galaxy Sample \citep[MaNDala;][]{CanoD2022}. Whereas MaNDala 
selected 136 galaxies at random from those passing these 
criteria, leaving the low-mass regime sparsely sampled, we 
apply the criteria to the full survey to obtain a 
substantially larger and more complete census. For each 
galaxy we extract elliptically binned, inverse variance 
weighted stacked spectra in up to ten radial bins, fit the 
H$\alpha$, [\ion{N}{ii}], H$\beta$, and [\ion{O}{iii}] 
emission lines simultaneously using Gaussian profiles, and 
derive gas-phase metallicity using the N2 calibration of 
\citet{PettiniPagel2004}. We characterise the radial 
metallicity profiles of the sample and investigate the 
dependence of metallicity and its radial gradient on local 
stellar mass surface density, light concentration, gas 
velocity dispersion, star formation rate, and large scale 
environment, to identify the primary physical drivers of 
chemical enrichment in this mass regime. This paper is 
organised as follows. Section~\ref{sec:data} describes the 
data and sample selection. Section~\ref{sec:pipeline} 
describes the analysis pipeline. Section~\ref{sec:results} 
presents the results. Section~\ref{sec:discussion} discusses 
the physical interpretation. Section~\ref{sec:conclusions} 
summarises our conclusions.

\section{Data and Sample}
\label{sec:data}
\subsection{The MaNGA Survey}
\label{sec:manga}
The Mapping Nearby Galaxies at Apache Point 
Observatory \citep[MaNGA;][]{Bundy2015} 
survey is one of the primary programs of 
the Sloan Digital Sky Survey IV 
\citep[SDSS-IV;][]{Blanton2017}. MaNGA 
used integral field spectroscopy to observe 
over 10,000 nearby galaxies with a 
dedicated 2.5\,m telescope at the Apache 
Point Observatory \citep{Gunn2006}, using 
hexagonal fiber bundle integral field units 
of 19 to 127 fibers, each fiber subtending 
$2\arcsec$ \citep{Drory2015}. The survey 
provides spectral coverage from $3600$ to 
$10300$\,\AA\ at resolution $R \sim 2000$ 
\citep{Smee2013}, with each galaxy observed 
out to 1.5 or 2.5 effective radii 
\citep{Law2015}. The data products used 
here are the LOGCUBE files from the MaNGA 
Data Reduction Pipeline 
\citep[DRP;][]{Law2016}, specifically the 
HYB10-MILESHC-MASTARHC2 version from Data 
Release 17 \citep{Abdurrouf2022}. Each 
LOGCUBE file contains the observed flux 
cube (\texttt{FLUX}) in units of 
$10^{-17}$\,erg\,s$^{-1}$\,cm$^{-2}$\,\AA$^{-1}$ 
per spaxel, the inverse variance 
(\texttt{IVAR}, the reciprocal of the 
variance per spaxel used to weight spectral 
fits), a bitmask (\texttt{MASK}, where a 
value of zero indicates a valid 
uncontaminated spaxel), a log-spaced 
wavelength array (\texttt{WAVE}), and the 
best-fit stellar continuum 
(\texttt{STELLAR}) from the MaNGA Data 
Analysis Pipeline 
\citep[DAP;][]{Westfall2019}, with a 
spaxel scale of $0.5\arcsec$.

\subsection{Sample Selection}
\label{sec:sample}
We select the largest sample of dwarf 
galaxies to date from Data Release 17 of 
the MaNGA survey. We start from the full 
MaNGA DR17 sample of 11,273 galaxies and 
apply a series of quality, redshift, mass, 
and resolution cuts to define a sample 
suitable for spatially resolved metallicity 
analysis. We first require a clean data 
reduction flag (\texttt{drp3qual} $= 0$), 
which reduces the sample to 7,229 galaxies. 
We then restrict the redshift range to 
$0.01 \leq z \leq 0.05$ and adopt a 
stellar mass cut 
$\log(M_{\star}/M_{\odot}) < 9.1$ from 
the NSA catalogue, leaving 4,961 and then 
480 galaxies respectively. We also exclude 
systems more luminous than the Large 
Magellanic Cloud with an absolute magnitude 
cut $M_{g} > -18.5$, which leaves 460 
galaxies. These criteria are consistent 
with those of the MaNGA Dwarf Galaxy Sample 
\citep[MaNDala;][]{CanoD2022}.

We then apply a spatial resolution cut, 
requiring $R_{\rm e} \geq \theta_{\rm PSF}$, 
where $\theta_{\rm PSF} = 2.5\arcsec$ is 
the MaNGA point spread function full width 
at half maximum. This cut excludes 
spatially unresolved galaxies, whose 
gradients would be biased toward positive 
values by beam smearing \citep{Acharyya2020}. 
We show in Appendix~\ref{app:psf} that 
even within the resolved sample a residual 
dependence of the measured gradient on 
$R_{\rm e}/\theta_{\rm PSF}$ persists, and 
we demonstrate that our main results are 
robust to applying a stricter cut of 
$R_{\rm e} \geq 2\,\theta_{\rm PSF}$. We 
also remove two galaxies with unphysical 
effective radii $> 50\arcsec$ from failed 
photometric fits. These steps leave 430 
galaxies. We then run the full stacking and 
emission line fitting pipeline on these 
galaxies and keep only those with at least 
three valid radial bins, which we define by 
$S/N > 3$ in both H$\alpha$ and 
[\ion{N}{ii}]. This removes 42 passive 
galaxies with no detectable emission and 6 
galaxies with fewer than three valid bins. 
Our final sample contains 382 star-forming 
low-mass galaxies.

We apply the same selection criteria to 
the full DR17 sample rather than to a 
random subset. Our sample therefore 
includes essentially all MaNDala galaxies 
that pass our quality, resolution, and 
emission line cuts, along with many 
additional low-mass galaxies from the rest 
of the survey.

\begin{figure*}
\centering
\includegraphics[width=\hsize]{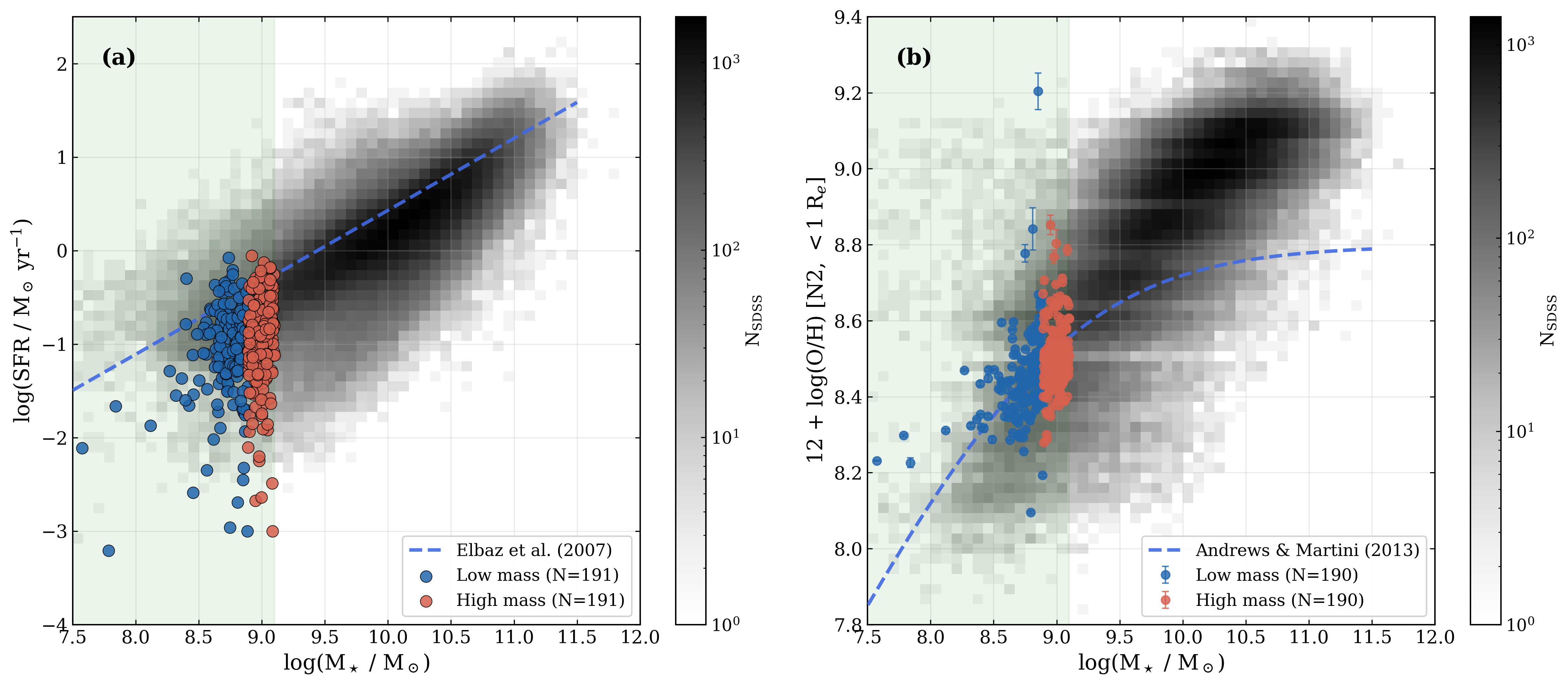}
\caption{Sample characterisation of the 
382 dwarf galaxies used in this work. 
\textit{Left panel:} star-forming main 
sequence, with our galaxies colour coded 
by stellar mass group relative to the 
sample median 
($\log(M_{\star}/M_{\odot}) = 8.89$). 
The grey histogram shows SDSS DR8 
star-forming galaxies from the MPA-JHU 
catalogue, and the dashed line is the 
star-forming main sequence of 
\citet{Elbaz2007}. Our galaxies lie 
systematically below this relation 
because it is calibrated for more massive 
galaxies; at low stellar masses the 
relation has a lower normalisation at 
fixed stellar mass. \textit{Right panel:} 
mass-metallicity relation, with our 
galaxies coloured by mass group over the 
SDSS star-forming population (grey), and 
the dashed line showing the 
mass-metallicity relation of 
\citet{AndrewsMartini2013}. The SDSS 
distribution and the 
\citet{AndrewsMartini2013} relation are 
shown on their native abundance scales, 
which differ from the N2 calibration used 
for our galaxies. Because the N2 
calibration saturates at higher 
metallicity, our points and the SDSS 
distribution are not directly comparable 
in absolute abundance at the high-mass 
end; the comparison shows the location of 
our sample in the mass-metallicity plane 
rather than an exact abundance match. The 
right panel shows 190 galaxies per group 
rather than 191 because one galaxy 
(8934-9102) has an unphysically high N2 
metallicity ($12+\log(\mathrm{O/H}) = 
9.20$) and is excluded from this panel 
but retained in the gradient analysis. 
The green shaded band marks the stellar 
mass range of our selection. SDSS 
background metallicities are central 
fibre values, while ours are inverse 
variance weighted means within 
$1\,R_{\rm e}$.}
\label{fig:sample}
\end{figure*}

The final sample spans stellar masses 
$7.2 \leq \log(M_{\star}/M_{\odot}) \leq 
9.1$. The redshift selection criterion 
was $0.01 \leq z \leq 0.05$. Three 
galaxies in the final sample 
(11020-12701, 8554-12704, and 8150-3701) 
have catalogue redshifts below this 
limit, in the range $z = 0.0002$ to 
$0.0017$, and were retained in error 
during the selection. We have verified 
that these three galaxies pass all 
other quality, resolution, and 
emission-line criteria and have valid 
gradient measurements, and that their 
removal leaves the concentration 
correlation ($r = 0.304$), the median 
gradient slope 
($-0.007$\,dex\,$R_{\rm e}^{-1}$), and 
the environmental null result unchanged. 
We therefore retain them. They are 
excluded from the $\Sigma_{\star}$ 
analysis in any case 
(Section~\ref{sec:sigma}), because 
peculiar velocities dominate the Hubble 
flow at these redshifts. In addition, 
for 23 galaxies we adopt 
H$\alpha$-derived redshifts 
(Section~\ref{sec:redshift}); these 
corrections are small and do not move 
any galaxy outside the selection range. 
The effective radii range from $2.5$ to 
$32.7\arcsec$, and $R_{\rm e}/\theta_{\rm 
PSF}$ ranges from $1.1$ to $13.1$. This 
is the largest spatially resolved 
spectroscopic sample of dwarf galaxies 
studied with MaNGA to date. For our 
comparative analysis we split the sample 
at its median stellar mass, 
$\log(M_{\star}/M_{\odot}) = 8.89$, into a 
lower mass group 
($\log(M_{\star}/M_{\odot}) < 8.89$, 
$N = 191$) and a higher mass group 
($\log(M_{\star}/M_{\odot}) \geq 8.89$, 
$N = 191$). The low-mass end of our sample 
remains limited because MaNGA was not 
designed to target very low-mass galaxies, 
which is a fundamental limitation of the 
survey rather than of our selection. We 
show the sample characterisation in 
Figure~\ref{fig:sample} and list the 
complete properties in 
Table~\ref{tab:master}.

Stellar masses are drawn from the 
NASA-Sloan Atlas 
\citep[NSA;][]{Blanton2011} and assume a 
\citet{Chabrier2003} initial mass function. 
We use the structural parameters, the 
effective radius $R_{\rm e}$, position 
angle, and axis ratio in the $r$ band, to 
define the elliptical geometry of our 
radial binning. We take these parameters 
from the Sérsic profile analysis of DESI 
Legacy Survey imaging \citep{CanoD2022} 
where available, and from the NSA catalogue 
and \texttt{drpall} file otherwise.

\subsection{Ancillary Catalogues}
\label{sec:ancillary}
We make use of several ancillary catalogues 
to characterise the physical properties and 
environments of our sample galaxies.

We derive integrated stellar population 
properties using the PIPE3D spectral 
fitting code \citep{Sanchez2016}. From 
PIPE3D we take the total star formation 
rate, the specific star formation rate, the 
luminosity- and mass-weighted stellar ages, 
the D4000 break strength, and the H$\alpha$ 
equivalent width. We derive the star 
formation rate from the H$\alpha$ emission 
line luminosity integrated over the full 
MaNGA field of view, using the 
\citet{Kennicutt1998} conversion corrected 
to a \citet{Chabrier2003} initial mass 
function. For galaxies without a PIPE3D 
measurement, we compute the star formation 
rate directly from the total H$\alpha$ flux 
in the \texttt{EMLINE\_GFLUX} extension of 
the DAP MAPS files. We integrate all 
unmasked spaxels with positive H$\alpha$ 
flux, convert to luminosity at each 
galaxy's redshift, and apply the same 
calibration. The two approaches give 
consistent results. No dust correction is 
applied to the H$\alpha$ fluxes used for 
star formation rate estimation.

We take morphological parameters from the 
MaNGA Visual Morphology Catalogue 
\citep{VazquezMata2022}, which provides 
visual classifications and non-parametric 
measurements from DESI Legacy Survey 
imaging. From this catalogue we use light 
concentration $C$, asymmetry $A$, 
morphological T-type, an edge-on flag, and 
a tidal feature flag. The catalogue does 
not cover the full MaNGA sample, so 
concentration is available for 230 of our 
galaxies. Of these 230 galaxies, 219 pass the 
gradient quality cut 
($\sigma_{\nabla\Delta Z} \leq 
0.05$\,dex\,$R_{\rm e}^{-1}$) and are 
used in the correlation analysis. The remaining galaxies have no 
entry, which is a coverage limitation of 
the external catalogue rather than a 
quality cut on our part. We find that the 
galaxies with a measured concentration are 
statistically indistinguishable from those 
without in stellar mass, effective radius, 
redshift, star formation rate, and gradient 
slope (Mann-Whitney $p > 0.1$ in every 
case). The coverage fraction is also 
roughly uniform with stellar mass, so the 
concentration subsample is not 
significantly biased.

We draw large scale environment parameters 
from the Galaxy Environment for MaNGA 
catalogue \citep[GEMA;][]{ArgudoFernandez2015}, 
which characterises the surroundings of 
MaNGA galaxies in the SDSS spectroscopic 
survey. From this catalogue we use the 
number of neighbours within 1 and 5\,Mpc, 
the group size, the group membership, the 
brightest group galaxy flag, the halo mass, 
and the isolation flag.

We measure the gas velocity dispersion from 
the \texttt{EMLINE\_GSIGMA} extension of 
the DAP MAPS files \citep{Westfall2019}, 
which gives the Gaussian fit velocity 
dispersion of each emission line per 
spaxel. We adopt the median H$\alpha$ 
velocity dispersion within $1\,R_{\rm e}$ 
as our tracer of turbulent gas motions, 
which we denote $\sigma_{\rm gas}$. This 
is available for all our galaxies.

\section{Analysis Pipeline}
\label{sec:pipeline}

\subsection{Continuum Subtraction}
\label{sec:continuum}
\begin{figure*}
\centering
\includegraphics[width=\hsize]{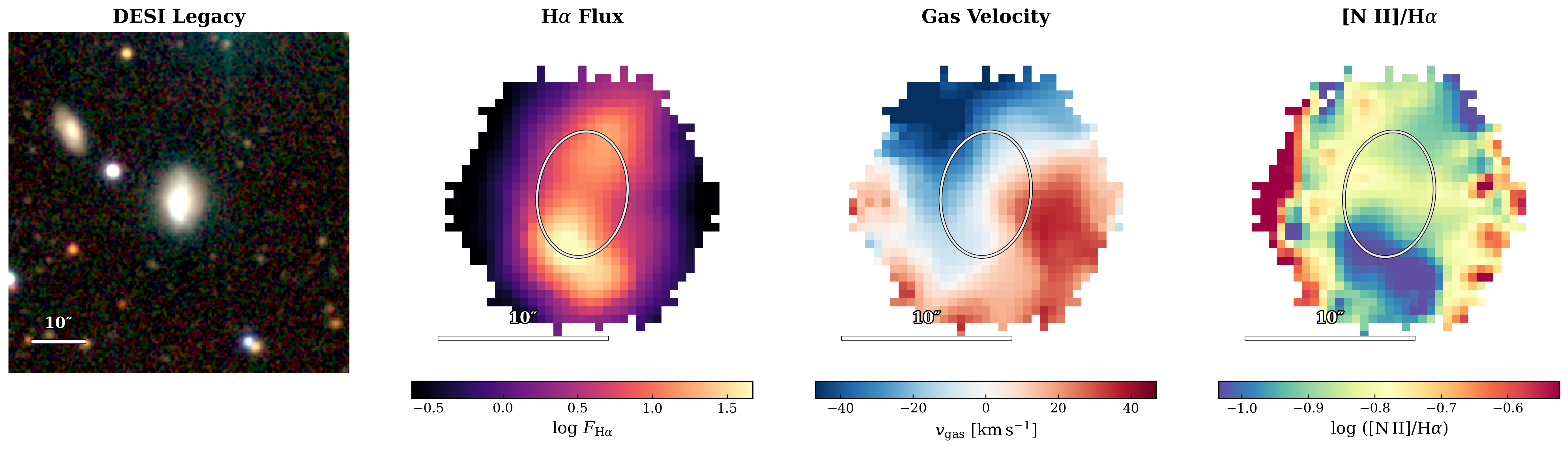}
\caption{A representative galaxy in our 
sample (MaNGA \texttt{Plate-IFU:} 
8563-3704, 
$\log(M_{\star}/M_{\odot}) = 8.63$). 
From left to right: DESI Legacy Survey 
RGB image, H$\alpha$ flux, gas velocity 
field, and [\ion{N}{ii}]/H$\alpha$ ratio, 
the latter being the N2 metallicity 
calibrator. The $1\,R_{\rm e}$ ellipse is 
marked on the map panels and $10\arcsec$ 
scale bars are shown. The velocity and 
[\ion{N}{ii}]/H$\alpha$ maps are 
restricted to spaxels with H$\alpha$ 
$S/N > 3$.}
\label{fig:maps_obs}
\end{figure*}

Before fitting the emission lines, we 
subtract the stellar continuum from each 
spaxel. We use the \texttt{STELLAR} 
extension of the LOGCUBE file, which is 
the best-fit stellar population model from 
the MaNGA Data Analysis Pipeline 
\citep{Westfall2019}, derived by fitting 
a linear combination of MILES stellar 
population templates \citep{Sanchez2006} 
to the observed spectrum using penalised 
pixel fitting \citep[pPXF;][]{Cappellari2017}. 
This leaves a residual spectrum that 
contains only the emission line signal and 
noise,
\begin{equation}
F_{\rm clean}(\lambda, x, y) = 
F_{\rm obs}(\lambda, x, y) 
- F_{\rm stellar}(\lambda, x, y),
\end{equation}
where $F_{\rm obs}$ is the observed flux 
from the \texttt{FLUX} extension. 
Figure~\ref{fig:maps_obs} shows the 
observed maps for a representative galaxy.

\subsection{Redshift Estimation}
\label{sec:redshift}
We adopt the systemic redshift of each 
galaxy from the NSA catalogue. As a 
consistency check, we also measure the 
redshift from the observed peak wavelength 
of H$\alpha$ in the central-bin stacked 
spectrum, which is the strongest line in 
our spectra. For the large majority of 
galaxies the two values agree to within 
the noise, and we retain the NSA redshift. 
For 23 galaxies the H$\alpha$-derived 
redshift differs from the NSA value by 
more than $100$\,km\,s$^{-1}$, and we 
adopt the H$\alpha$ redshift instead, as 
it is measured directly from the spectral 
data. For 22 of these galaxies the 
velocity difference is between $100$ and 
$130$\,km\,s$^{-1}$, consistent with 
small systematic offsets in the NSA 
photometric redshifts at low redshift. 
One galaxy (8719-9101) shows a larger 
discrepancy of $2754$\,km\,s$^{-1}$, 
which we attribute to an incorrect NSA 
redshift assignment, likely from a 
neighbouring brighter object. For this galaxy the emission-line 
fitting was performed within a 
$\pm400$\,km\,s$^{-1}$ window centred 
on the NSA redshift, so the measured 
gradient reflects the spectral features 
at that redshift position regardless of 
the adopted systemic value. The gradient 
measurement for this galaxy should 
therefore be treated with caution. In all 
cases the emission line measurements 
themselves are unaffected by this 
correction, as the redshift enters only 
through the rest-frame wavelength 
calibration and not through the flux or 
signal-to-noise of the stacked spectra. 
As a result of these corrections, a 
small number of galaxies in the final 
sample have adopted redshifts below the 
nominal lower limit of $z = 0.01$. The 
adopted redshift defines the rest-frame 
wavelength array $\lambda_{\rm rest} = 
\lambda_{\rm obs}/(1+z)$ that we use to 
fit all four emission lines.

\subsection{Spatial Geometry and Elliptical Binning}
\label{sec:binning}

For each galaxy we define elliptical radial bins using the 
position angle and axis ratio described in 
Section~\ref{sec:sample}. The elliptical radius of each 
spaxel is
\begin{equation}
r_{\rm ellip} = \sqrt{x_{\rm rot}^{2} + 
\left(\frac{y_{\rm rot}}{b/a}\right)^{2}} 
\times 0.5\arcsec,
\end{equation}
where $x_{\rm rot}$ and $y_{\rm rot}$ are the spaxel 
coordinates rotated to align with the major axis, $b/a$ is 
the axis ratio, and the factor $0.5\arcsec$ converts spaxels 
to arcseconds. The rotation is
\begin{equation}
x_{\rm rot} = x\cos\phi + y\sin\phi, \qquad
y_{\rm rot} = -x\sin\phi + y\cos\phi,
\end{equation}
where $\phi = \phi_{\rm PA} - 90^{\circ}$. Here 
$\phi_{\rm PA}$ is the galaxy position angle from the 
structural parameters (Section~\ref{sec:sample}), which we 
convert to the mathematical convention. We normalise the 
radius as $r_{\rm norm} = r_{\rm ellip}/R_{\rm e}$.

\begin{figure*}
\centering
\includegraphics[width=\hsize]{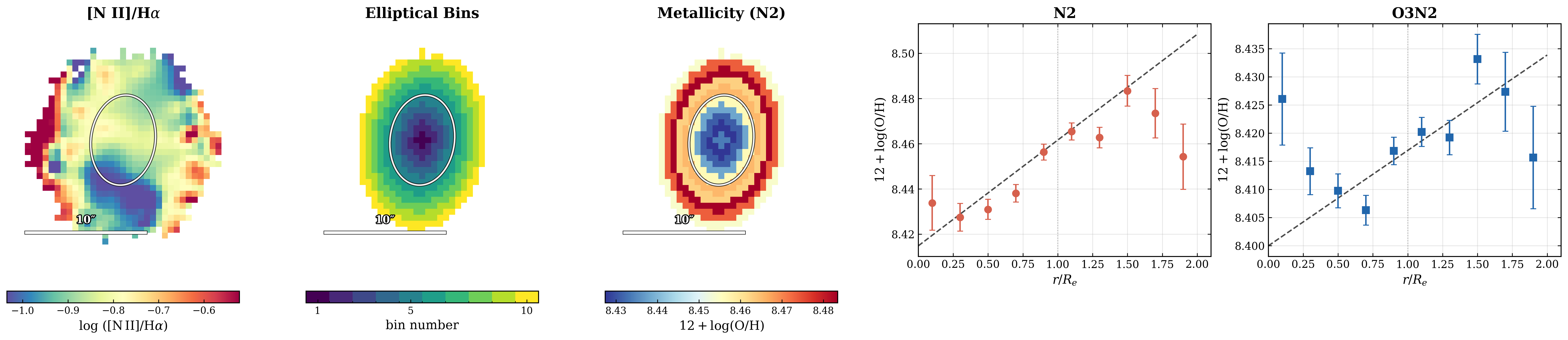}
\caption{Analysis sequence for the 
representative galaxy 8563-3704. From 
left to right: [\ion{N}{ii}]/H$\alpha$ 
ratio map, elliptical binning map 
(0.2\,$R_{\rm e}$ bins to $2\,R_{\rm e}$), 
two-dimensional N2 metallicity map, and 
the radial metallicity profiles derived 
using the N2 (fourth panel) and O3N2 
(fifth panel) calibrations of 
\citet{PettiniPagel2004}. Placing the 
[\ion{N}{ii}]/H$\alpha$ map beside the 
N2 metallicity map shows directly how 
the line ratio translates into 
$12 + \log(\mathrm{O/H})$. For this 
galaxy the N2 and O3N2 profiles yield 
consistent radial slopes; a statistical 
comparison of both calibrations across 
the full sample is presented in 
Appendix~\ref{app:calibration}.}
\label{fig:maps}
\end{figure*}

We define ten annular bins of width $0.2\,R_{\rm e}$ from 
$0$ to $2.0\,R_{\rm e}$, with centres at $0.1, 0.3, \dots, 
1.9\,R_{\rm e}$. Figure~\ref{fig:maps} shows the binning 
scheme, the resulting metallicity map, and the radial 
profiles for a representative galaxy.

When we compute median radial profiles, we take the median 
at each bin centre over only those galaxies with a valid 
measurement at that radius. The outer bins of the median 
profile therefore rest on fewer galaxies than the inner 
bins, and are correspondingly less certain in 
Figure~\ref{fig:gradients}.

\subsection{Spectral Stacking}
\label{sec:stacking}

Within each bin we combine the continuum subtracted spectra 
of all valid spaxels using an inverse variance weighted mean,
\begin{equation}
F_{\rm stack}(\lambda) = \frac{\sum_{i} 
F_{i}(\lambda)\, w_{i}(\lambda)}{\sum_{i} w_{i}(\lambda)},
\end{equation}
where $w_{i}(\lambda) = \sigma_{i}^{-2}(\lambda)$ is the 
inverse variance weight from the \texttt{IVAR} extension. We 
include only spaxels with \texttt{MASK}$= 0$ near H$\alpha$, 
and we discard bins with fewer than three valid spaxels. We 
propagate the stacked inverse variance, 
$w_{\rm stack}(\lambda) = \sum_{i} w_{i}(\lambda)$, into the 
line flux uncertainties.

\subsection{Emission Line Fitting}
\label{sec:fitting}
We fit each stacked spectrum with two 
simultaneous multi-component Gaussian 
models, one for the 
H$\alpha$\,+\,[\ion{N}{ii}] complex and 
one for the 
H$\beta$\,+\,[\ion{O}{iii}] complex 
\citep{Belfiore2017}.

For the H$\alpha$\,+\,[\ion{N}{ii}] 
complex we fit H$\alpha$ 
($6562.8$\,\AA), 
[\ion{N}{ii}]$\lambda6583$ 
($6583.4$\,\AA), and 
[\ion{N}{ii}]$\lambda6548$ 
($6548.1$\,\AA). All three components 
share one velocity offset and width 
$\sigma$, and we fix the 
[\ion{N}{ii}]$\lambda6548$/$\lambda6583 
= 1/3$ ratio at its theoretical value 
\citep{StoreyZeippen2000}, since the two 
lines share the same upper energy level. 
This leaves four free parameters, the 
two amplitudes, the offset, and $\sigma$. 
We constrain the offset to 
$\pm400$\,km\,s$^{-1}$ and $\sigma$ to 
$0.5$ to $8.0$\,\AA. We fit the 
H$\beta$\,+\,[\ion{O}{iii}] complex in 
the same way for H$\beta$ 
($4861.3$\,\AA), 
[\ion{O}{iii}]$\lambda5007$ 
($5006.8$\,\AA), and 
[\ion{O}{iii}]$\lambda4959$ 
($4958.9$\,\AA), fixing the 
[\ion{O}{iii}]$\lambda4959$/$\lambda5007 
= 1/2.98$ ratio 
\citep{StoreyZeippen2000} and using the 
same offset and $\sigma$ constraints.

We retain a bin only if both fits 
converge, $\sigma$ lies within bounds, 
all amplitudes are positive, and 
$S/N \geq 3$ in both H$\alpha$ and 
[\ion{N}{ii}]$\lambda6583$. The N2 
calibration depends only on these two 
lines, so we impose no $S/N$ requirement 
on H$\beta$ or [\ion{O}{iii}]$\lambda5007$ 
for the bins used in N2. Where we also 
compute the O3N2 calibration, we 
additionally require $S/N \geq 3$ in 
both of those lines. This relaxed N2 
criterion recovers additional valid bins 
in the low surface brightness outer 
regions, where H$\beta$ and 
[\ion{O}{iii}] are too faint to detect. 
We define the signal-to-noise ratio as
\begin{equation}
S/N = \frac{F_{\rm line}}{\sigma_{F}},
\end{equation}
where the flux error is
\begin{equation}
\sigma_{F} = \sqrt{\sum_{j} 
w_{\rm stack}^{-1}(\lambda_{j})} \cdot 
\sigma \cdot \sqrt{2\pi},
\end{equation}
summed over pixels within $3\sigma$ of 
the line centre, where $\sigma$ is the 
fitted Gaussian width in \AA.

\subsection{Metallicity Calibration}
\label{sec:metallicity}

We derive gas-phase metallicity using the 
strong line calibrations of 
\citet{PettiniPagel2004}. Throughout this 
paper we use the symbol $Z$ as a shorthand 
for the logarithmic oxygen abundance 
$12+\log(\mathrm{O/H})$, not the total 
metal mass fraction. Our primary indicator 
is the N2 index,
\begin{equation}
12 + \log(\mathrm{O/H}) = 8.90 + 0.57 
\times \log\left(\frac{[\ion{N}{ii}]
\lambda6583}{\mathrm{H}\alpha}\right).
\end{equation}
We also use the O3N2 index of 
\citet{PettiniPagel2004} to check how 
much our results change relative to the 
N2 calibration. We define it as
\begin{equation}
\mathrm{O3N2} = \log\left(
\frac{[\ion{O}{iii}]\lambda5007/
\mathrm{H}\beta}{[\ion{N}{ii}]\lambda6583/
\mathrm{H}\alpha}\right),
\end{equation}
which gives
\begin{equation}
12 + \log(\mathrm{O/H}) = 8.73 - 0.32 
\times \mathrm{O3N2}.
\end{equation}
We adopt N2 as our primary calibration 
for two reasons. First, it uses only 
H$\alpha$ and [\ion{N}{ii}], which are 
closely spaced in wavelength and so 
insensitive to differential extinction 
and flux calibration uncertainties 
\citep{PettiniPagel2004}. We can also 
measure N2 in the outer low surface 
brightness bins, where H$\beta$ and 
[\ion{O}{iii}] are undetected. Second, 
we show in Figure~\ref{fig:maps} that 
adopting O3N2 instead leaves the shape 
and slope of the radial metallicity 
gradient essentially unchanged for the 
representative galaxy shown. A full 
statistical comparison of both 
calibrations across the sample is 
presented in Appendix~\ref{app:calibration}. 
N2 is also widely used in the literature, 
which makes our gradients straightforward 
to compare with previous work.

We propagate the metallicity uncertainties 
analytically from the line flux errors. 
For N2,
\begin{equation}
\sigma_{Z,\rm N2} = \frac{0.57}{\ln 10} 
\sqrt{\left(\frac{\sigma_{\rm NII}}
{F_{\rm NII}}\right)^{2} + 
\left(\frac{\sigma_{\rm H\alpha}}
{F_{\rm H\alpha}}\right)^{2}},
\end{equation}
and for O3N2,
\begin{equation}
\sigma_{Z,\rm O3N2} = \frac{0.32}{\ln 10} 
\sqrt{\sum_{X} \left(\frac{\sigma_{X}}
{F_{X}}\right)^{2}},
\end{equation}
summed over $X \in \{$[\ion{O}{iii}]
$\lambda5007$, H$\beta$, 
[\ion{N}{ii}]$\lambda6583$, 
H$\alpha\}$.

\subsection{Stellar Mass Surface Density}
\label{sec:sigma}

For each galaxy we compute the stellar 
mass surface density $\Sigma_{\star}$ per 
spaxel. We integrate the best-fit stellar 
continuum (\texttt{STELLAR}) over the 
rest-frame range $5000$ to $7000$\,\AA\ 
to obtain a luminosity proxy map, which 
we normalise by the total integrated 
luminosity to give each spaxel's 
fractional contribution to the stellar 
light,
\begin{equation}
\Sigma_{\star}(x, y) = M_{\star} \times 
\frac{L(x,y)}{L_{\rm total}} \times 
\frac{1}{A_{\rm spaxel}},
\end{equation}
where $M_{\star}$ is the total stellar 
mass from the NSA catalogue, 
$L(x,y)/L_{\rm total}$ is the fractional 
contribution of each spaxel to the 
integrated stellar light, and 
$A_{\rm spaxel}$ is the physical area per 
spaxel in kpc$^{2}$. This approach 
assumes a spatially constant 
mass-to-light ratio across all spaxels. 
While per-spaxel stellar mass maps from 
full spectral fitting would provide a 
more direct estimate of $\Sigma_{\star}$, 
we expect this approximation to have a minor effect on our results, since 
low-mass star-forming galaxies tend to 
have relatively uniform stellar populations 
across their discs, limiting the magnitude 
of any radial mass-to-light ratio gradient. We compute 
$A_{\rm spaxel}$ from the angular 
diameter distance, assuming a flat 
$\Lambda$CDM cosmology 
($H_{0} = 70$\,km\,s$^{-1}$\,Mpc$^{-1}$, 
$\Omega_{\rm M} = 0.3$, 
$\Omega_{\Lambda} = 0.7$). The resulting 
$\Sigma_{\star}$ has units of 
$M_{\odot}\,\mathrm{kpc}^{-2}$, and we 
use $\log(\Sigma_{\star})$ throughout. 
For each radial bin we adopt the median 
$\Sigma_{\star}$ as its representative 
value. We exclude three galaxies with 
$z < 0.002$ here, because peculiar 
velocities dominate over the Hubble flow 
at such low redshifts and make the 
angular diameter distance, and hence 
$A_{\rm spaxel}$, unreliable 
\citep{Blanton2005}. This leaves 379 
galaxies with $\Sigma_{\star}$.

\subsection{Residual Metallicity}
\label{sec:residual}

The metallicity of a galaxy depends 
primarily on its stellar mass, which is 
traced locally by the stellar mass surface 
density $\Sigma_{\star}$. We want to 
subtract out this primary dependence and 
recover any secondary dependencies on 
other physical parameters. We therefore 
define a residual metallicity $\Delta Z$ 
by subtracting from each bin the value 
predicted by a global linear fit between metallicity 
and $\log(\Sigma_{\star})$ across all 
valid bins, performed using ordinary 
least squares without weighting by 
individual measurement uncertainties, 
treating each bin as an independent 
data point,
\begin{equation}
\Delta Z = Z_{\rm N2} - 
Z_{\rm pred}(\Sigma_{\star}),
\end{equation}
where $Z_{\rm pred}(\Sigma_{\star}) = 
0.0963 \times \log(\Sigma_{\star}) + 
7.7643$. We fit this relation across all 
3608 valid bins from the full sample 
(Pearson $r = 0.371$, 
$p = 2.0 \times 10^{-118}$), and we show 
it in Figure~\ref{fig:zsigma} in 
Appendix~\ref{app:zsigma}. We also 
subtract a small systematic radial trend 
of $+0.017$\,dex\,$R_e^{-1}$ from the 
residual profiles to account for the 
non-zero mean gradient introduced by the 
linear approximation to the 
$Z$--$\Sigma_{\star}$ relation; we 
demonstrate in Appendix~\ref{app:nonlinear} 
that this correction leaves all 
correlation results unchanged. A positive 
$\Delta Z$ means a bin is more metal rich 
than its local stellar mass surface 
density predicts, and a negative value 
means it is more metal poor.

\subsection{Gradient Slope Fitting}
\label{sec:gradient}

For each galaxy we measure two gradient slopes by fitting 
straight lines to its radial profiles with weighted least 
squares ($\chi^{2}$ minimisation), weighting each bin by the 
inverse variance of its metallicity. Fitting $Z_{\rm N2}$ 
against normalised radius $r/R_{\rm e}$ gives the metallicity 
gradient $\nabla Z$, and fitting the residual metallicity 
$\Delta Z$ in the same way gives the residual metallicity 
gradient $\nabla \Delta Z$. Both slopes are in units of 
dex\,$R_{\rm e}^{-1}$. We adopt the scaled slope uncertainty, 
which inflates the formal error by the residual scatter about 
the fit and so accounts for intrinsic profile scatter beyond 
the measurement errors. We include only galaxies with at 
least three valid bins.

For the correlation analysis (Section~\ref{sec:results}) we 
further restrict to a clean sample of 366 galaxies with a 
well-constrained gradient slope, 
$\sigma_{\nabla\Delta Z} \leq 0.05$\,dex\,$R_{\rm e}^{-1}$. We set 
this threshold by fit quality rather than by any outcome. It 
removes only 13 of 379 galaxies, and the excluded objects are 
demonstrably poorly constrained. They have a median of six 
valid radial bins, against ten for the retained sample, and a 
median slope uncertainty of $0.063$\,dex\,$R_{\rm e}^{-1}$, 
against $0.009$ for the clean sample. The cut therefore 
isolates unreliable, often non-linear profiles, including 
several confirmed mergers, independently of any correlation.
\subsection{Visual Environmental Classification}
\label{sec:visual_class}
To complement the GEMA environment 
parameters (Section~\ref{sec:ancillary}), 
we visually classified all our galaxies 
using DESI Legacy Survey DR10 colour 
images from the Legacy Survey viewer. For 
each galaxy we downloaded a cutout large 
enough to include the MaNGA field of view 
and its immediate surroundings, typically 
$60$--$120\arcsec$ on a side depending on 
the galaxy size, and generated an RGB 
image from the $g$, $r$, and $z$ bands 
with an arcsinh stretch \citep{Lupton2004}. 
We placed each galaxy into one of two 
classes. We classified a galaxy as 
\textit{isolated} when it showed no 
companions, tidal features, or signs of 
interaction in the field or wider 
surroundings. We classified it as 
\textit{disturbed} when it showed clear 
interaction signatures, such as close 
companions, tidal tails, asymmetric 
morphology, or a companion overlapping 
the integral field unit footprint.

The light concentration $C$ and asymmetry 
$A$ used in our analysis are taken from 
the MaNGA Visual Morphology Catalogue 
\citep{VazquezMata2022} and are measured 
from DESI Legacy Survey $r$-band imaging. 
Concentration is defined as 
$C = 5\log(r_{80}/r_{20})$, where $r_{80}$ 
and $r_{20}$ are the radii enclosing 80 
and 20 per cent of the total flux within 
the Petrosian ellipse \citep{Conselice2003}. 
Asymmetry is defined as 
$A = \sum |I - I_{180}| / \sum |I|$, 
where $I_{180}$ is the galaxy image 
rotated by $180^{\circ}$ about its centre 
\citep{Abraham1996}.

This classification yielded 182 isolated 
and 200 disturbed galaxies. As an 
independent check, we compared the 
asymmetry of the two classes. Disturbed 
galaxies have a higher mean asymmetry 
($A = 0.291$) than isolated galaxies 
($A = 0.180$), and the difference is 
significant (Mann-Whitney $p = 10^{-4}$), 
which validates the classification. We 
record the classification flag in the 
master catalogue and use it in the 
environmental analysis 
(Section~\ref{sec:correlations}).

\section{Results}
\label{sec:results}

\subsection{Sample Characterisation}
\label{sec:sample_char}
Our final sample lies on the star-forming 
main sequence, which confirms that these 
are actively star-forming systems 
(Figure~\ref{fig:sample}, left panel). The 
galaxies follow the low-mass extension of 
the relation of \citet{Elbaz2007}, though 
with a systematic offset toward lower star 
formation rates at fixed stellar mass, as 
expected for the low-mass regime 
(Section~\ref{sec:ancillary}). The right 
panel shows the mass-metallicity relation 
from the N2 calibration. We compare it to 
the SDSS DR8 background from the MPA-JHU 
catalogue and to the reference relation of 
\citet{AndrewsMartini2013}, noting that 
the SDSS distribution and the literature 
relation are shown on their native 
abundance scales which differ from our N2 
calibration and are therefore not directly 
comparable in absolute metallicity. Our 
sample shows a significant positive 
correlation between stellar mass and 
gas-phase metallicity (Pearson $r = 0.444$, 
$r^2 = 0.197$, $p < 0.001$), though 
stellar mass accounts for only 20 per cent 
of the observed variance in metallicity, 
indicating that other processes also 
contribute to the scatter. The SDSS 
background metallicities are single 
$3\arcsec$ fibre values, which represent 
roughly the central metallicity of each 
galaxy. Ours are inverse variance weighted 
means within $1\,R_{\rm e}$, which for 
galaxies with negative gradients lie 
slightly below the central value.

\subsection{Metallicity and Residual 
Metallicity Gradients}
\label{sec:gradients}
\begin{figure*}
\centering
\includegraphics[width=\hsize]{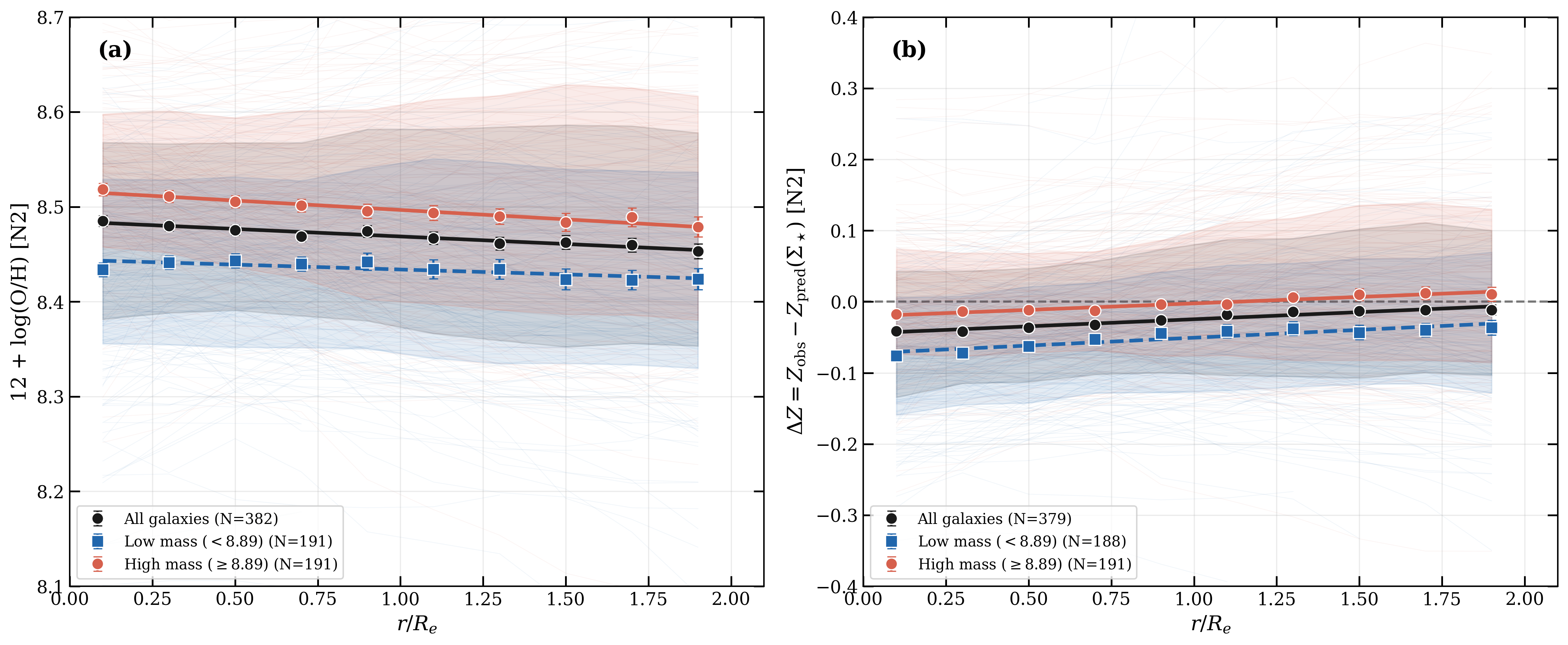}
\caption{\textit{Left:} radial gas-phase 
metallicity profiles for all 382 galaxies. 
Faint coloured lines are individual 
galaxies; markers show the median 
metallicity per radial bin for the higher 
mass ($\log(M_{\star}/M_{\odot}) \geq 
8.89$, red) and lower mass ($< 8.89$, 
blue) groups, split at the sample median, 
with error bars giving the standard error 
$\sigma/\sqrt{N}$ and shading the 16th 
to 84th percentile range. The solid and 
dashed straight lines are linear fits to 
the binned medians of the higher and lower 
mass groups respectively. \textit{Right:} 
residual metallicity 
$\Delta Z = Z_{\rm obs} - Z_{\rm pred}
(\Sigma_{\star})$ profiles for the same 
groups. Three galaxies with $z < 0.002$ 
are excluded from this panel because 
their $\Sigma_{\star}$ measurement is 
unreliable at such low redshifts 
(Section~\ref{sec:sigma}); symbols, 
errors, and fit lines as in the left 
panel. The horizontal dashed line marks 
$\Delta Z = 0$. The fit to the binned 
median profile is nearly flat in 
metallicity (slope 
$-0.016$\,dex\,$R_{\rm e}^{-1}$ for the 
full sample) and rises gently in the 
residual ($+0.020$\,dex\,$R_{\rm e}^{-1}$); 
these describe the plotted median trend 
and differ from the median of the 
per-galaxy gradient slopes quoted in the 
text. The higher mass group lies 
systematically above the lower mass group 
in both panels.}
\label{fig:gradients}
\end{figure*}

The left panel of Figure~\ref{fig:gradients} 
shows the radial metallicity profiles of 
our galaxies, with individual profiles in 
the background and median profiles for the 
low and high mass groups. We split the 
groups at the sample median 
$\log(M_{\star}/M_{\odot}) = 8.89$. We find 
that metallicity gradients are flat on 
average, with a median slope of 
$-0.007$\,dex\,$R_{\rm e}^{-1}$ and a 
standard deviation of 
$0.071$\,dex\,$R_{\rm e}^{-1}$. The 
distribution is strongly peaked about zero. 
Most galaxies (87 per cent) have slopes 
within $\pm 0.1$\,dex\,$R_{\rm e}^{-1}$, 
and only a handful of outliers reach 
$|\nabla Z| \gtrsim 0.2$\,dex\,$R_{\rm 
e}^{-1}$. This near-zero distribution is 
broadly consistent with efficient metal 
mixing in shallow gravitational potential 
wells \citep{Mercado2021}, though we note 
that beam smearing may contribute to the 
observed flattening in the less well 
resolved galaxies; we demonstrate in 
Appendix~\ref{app:psf} that the median 
slope becomes more negative 
($-0.026$\,dex\,$R_{\rm e}^{-1}$) when 
a stricter resolution cut is applied. Both 
mass groups have equally flat gradients, 
but the high mass group lies at higher 
metallicity than the low mass group at 
every radius. The two groups therefore 
differ in their overall metallicity but 
not in the shape of their gradients. More 
massive galaxies are uniformly more metal 
rich, while the way metallicity changes 
with radius is similar regardless of mass.

We quote two complementary measures of 
the gradient. The median slope above, 
$-0.007$\,dex\,$R_{\rm e}^{-1}$, comes 
from fitting a straight line to each 
galaxy individually and taking the median 
of the resulting per-galaxy slopes. This 
is a property of the galaxy population 
and has a well-defined distribution 
(standard deviation 
$0.071$\,dex\,$R_{\rm e}^{-1}$), and it 
is the quantity we use throughout our 
correlation analysis 
(Section~\ref{sec:correlations}). The 
straight lines drawn in 
Figure~\ref{fig:gradients} are different. 
We fit them to the binned median profile 
of each group, that is, to the stacked 
medians rather than to individual 
galaxies, and they have a slightly steeper 
slope ($-0.016$\,dex\,$R_{\rm e}^{-1}$ 
for the full sample). Both measures 
describe flat gradients. We adopt the 
per-galaxy median as our primary 
measurement because it retains the 
galaxy-to-galaxy distribution we need for 
the statistical analysis, and we use the 
binned-median fit only to guide the eye 
in the figure.

We find a statistically significant 
positive correlation between gas-phase 
metallicity and local stellar mass surface 
density across all valid bins 
($Z = 0.0963 \times \log(\Sigma_{\star}) 
+ 7.7643$, Pearson $r = 0.371$, 
$r^2 = 0.138$, $p = 2.0 \times 10^{-118}$), 
indicating that denser regions are on 
average more metal rich, though local 
stellar mass surface density accounts for 
only 14 per cent of the observed variance 
in metallicity. This relation holds across 
all radii and is not driven by any single 
radial zone. The two mass groups remain 
offset in metallicity even at fixed 
$\Sigma_{\star}$. This shows that total 
stellar mass affects metallicity 
independently of local density, and 
motivates the residual metallicity 
$\Delta Z$ that we define in 
Section~\ref{sec:residual}.

The right panel of 
Figure~\ref{fig:gradients} shows the 
$\Delta Z$ profiles, computed by removing 
the $\Sigma_{\star}$ dependence as 
described in Section~\ref{sec:residual}. 
A clear mass offset remains. The high 
mass group sits systematically above 
$\Delta Z = 0$ at most radii, while the 
low mass group sits below. This suggests 
that total stellar mass has an independent 
effect on chemical enrichment beyond what 
local stellar density predicts, and the 
offset persists at all radii.

\subsection{What Drives the Metallicity 
Gradient?}
\label{sec:correlations}

We now look for the physical properties 
that drive the galaxy-to-galaxy diversity 
in the metallicity gradient. Our main test 
is the correlation between each property 
and the residual gradient slope 
$\nabla \Delta Z$, which is the slope of 
a straight-line fit to the residual 
metallicity $\Delta Z$ against $r/R_{\rm e}$ 
(Sections~\ref{sec:residual} 
and~\ref{sec:gradient}). Since $\Delta Z$ 
already has the primary mass dependence 
removed, a correlation with 
$\nabla \Delta Z$ isolates the effect of 
each property on the radial metallicity 
structure, independent of stellar mass. 
We test stellar mass, star formation rate, 
gas velocity dispersion, light 
concentration, asymmetry, and the large 
scale environment from visual 
classification and the GEMA catalogue. 
We report both Pearson and Spearman 
correlation coefficients throughout, and 
we treat a correlation as significant when 
$p < 0.05$ in both tests. We also report 
the coefficient of determination $r^2$, 
which gives the fraction of variance in 
$\nabla \Delta Z$ explained by each 
property, as a measure of effect size.

\begin{figure*}
\centering
\includegraphics[width=\hsize]{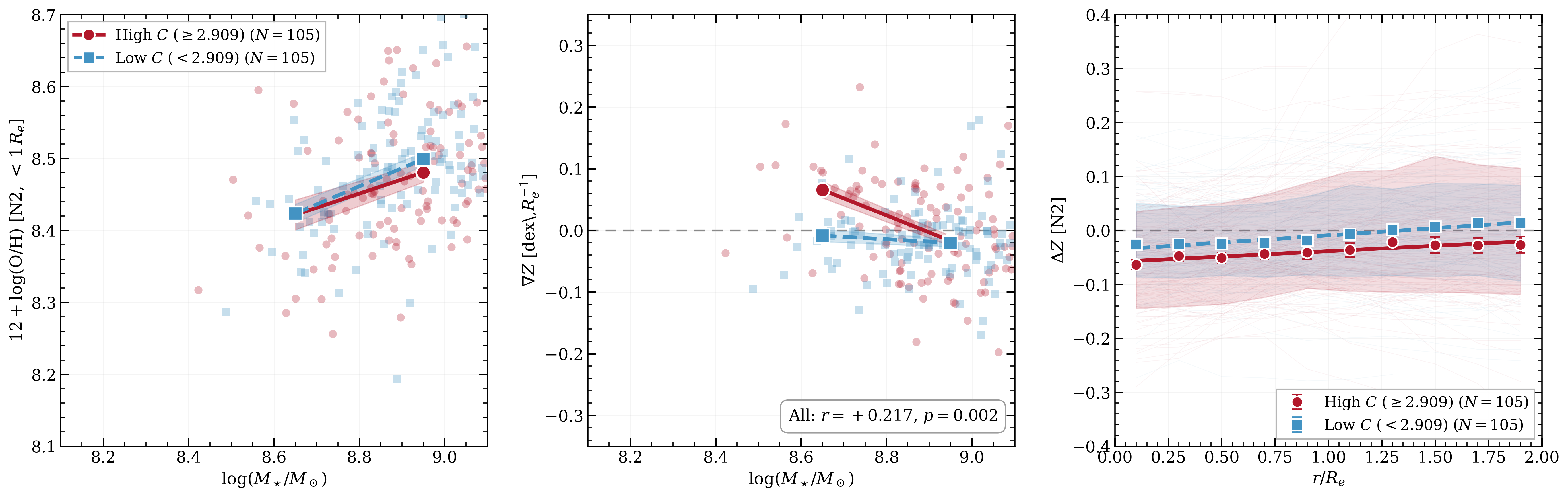}
\caption{Dependence of gas-phase 
metallicity on light concentration $C$, 
split at the median $C = 2.909$ into 
high-$C$ (red circles, solid) and low-$C$ 
(blue squares, dashed) mass-matched 
subsamples. \textit{Left:} inner 
metallicity $Z(1\,R_{\rm e})$, the inverse 
variance weighted mean N2 metallicity 
within $1\,R_{\rm e}$ 
(Section~\ref{sec:correlations}), against 
stellar mass. Small markers are individual 
galaxies and large connected markers are 
the median trend in $0.3$\,dex mass bins 
with standard errors. This panel recovers 
the mass-metallicity relation and shows 
whether the overall metallicity level 
depends on concentration. 
\textit{Centre:} raw gradient slope 
$\nabla Z$ (Section~\ref{sec:gradient}), 
from a straight-line fit to $Z_{\rm N2}$ 
against $r/R_{\rm e}$, against stellar 
mass. The dashed line marks a flat 
gradient, and the annotation gives the 
Pearson correlation between $C$ and 
$\nabla Z$ on the mass-matched 
subsamples. \textit{Right:} radial 
profiles of the residual metallicity 
$\Delta Z = Z_{\rm obs} - 
Z_{\rm pred}(\Sigma_{\star})$ 
(Section~\ref{sec:residual}). Faint lines 
are individual galaxies, and bold markers 
are the median per group with standard 
errors and a 16th to 84th percentile 
band. The dashed line marks $\Delta Z = 0$. 
The high-concentration subsample shows 
the clearest separation of residual 
profiles of any property we test.}
\label{fig:conc}
\end{figure*}

\begin{figure*}
\centering
\includegraphics[width=\hsize]{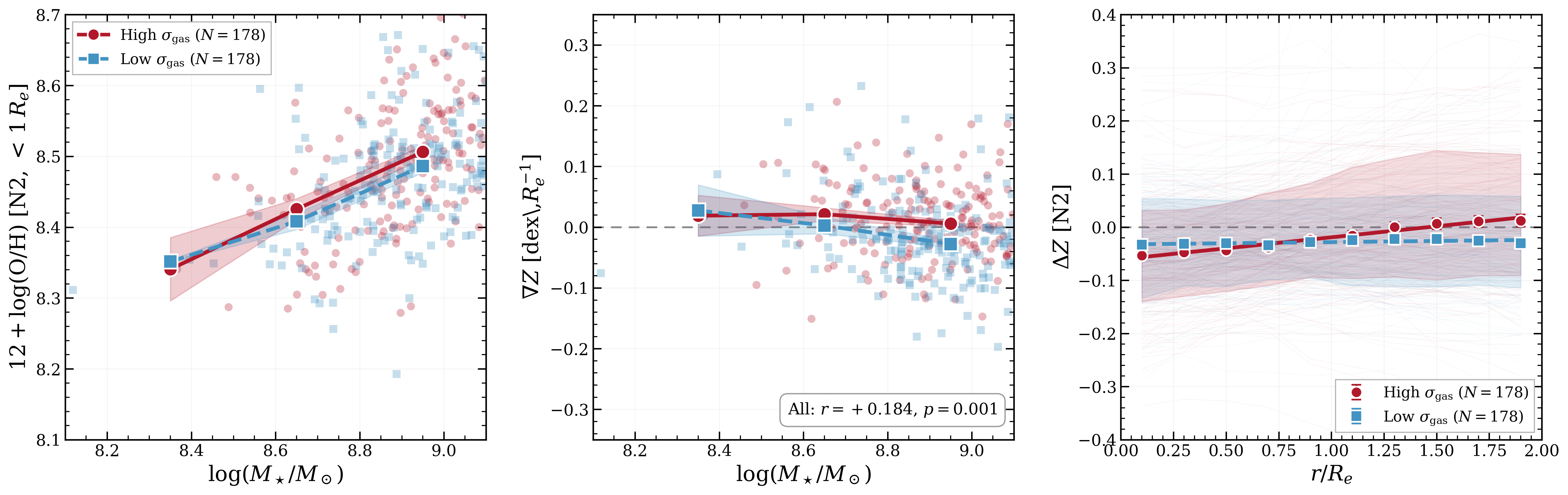}
\caption{As Figure~\ref{fig:conc}, but 
split at the median gas velocity 
dispersion 
$\sigma_{\rm gas} = 72.5$\,km\,s$^{-1}$ 
into high-$\sigma_{\rm gas}$ (red circles, 
solid) and low-$\sigma_{\rm gas}$ (blue 
squares, dashed) mass-matched subsamples. 
The left panel shows little dependence of 
the overall metallicity level on 
$\sigma_{\rm gas}$. The centre panel shows 
a positive correlation between 
$\sigma_{\rm gas}$ and the raw gradient 
slope on the mass-matched subsamples 
($r = +0.184$, $p = 0.001$).
The right panel shows that the residual 
profiles of the high-dispersion subsample 
lie systematically above those of the 
low-dispersion subsample. We note that 
$\sigma_{\rm gas}$ shows some dependence 
on spatial resolution and the result 
should be interpreted with caution.}
\label{fig:sigmagas}
\end{figure*}

\begin{figure*}
\centering
\includegraphics[width=\hsize]{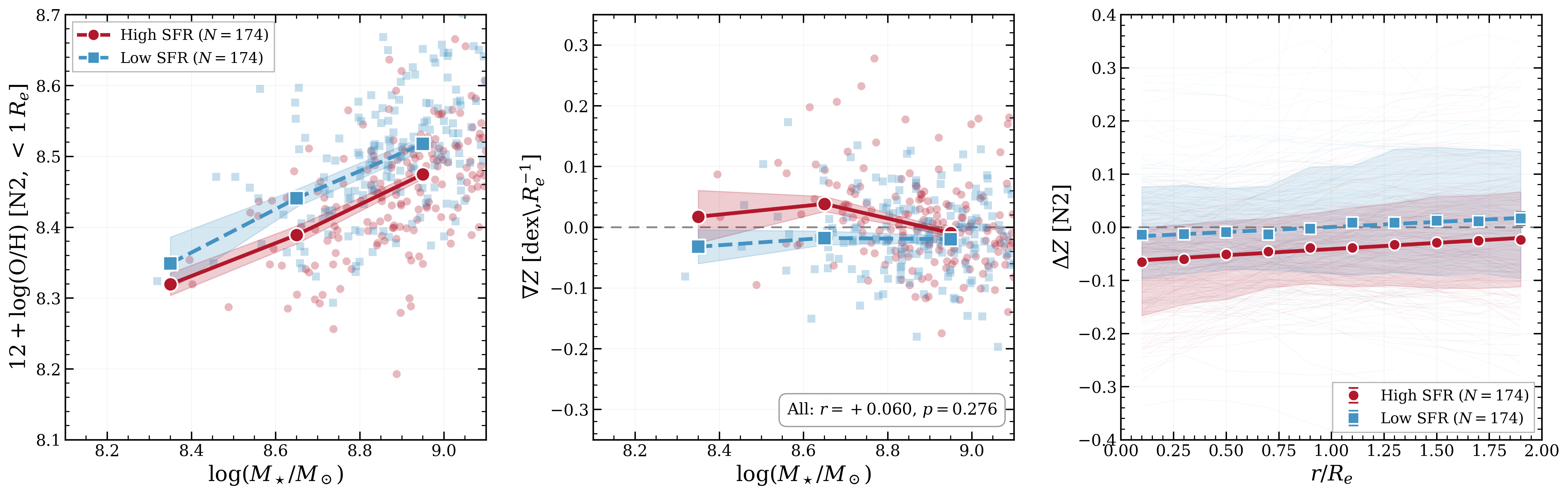}
\caption{As Figure~\ref{fig:conc}, but 
split at the median star formation rate 
$\log(\mathrm{SFR}) = -0.952$ into 
high-SFR (red circles, solid) and low-SFR 
(blue squares, dashed) mass-matched 
subsamples. The left panel recovers the 
mass-metallicity relation with no clear 
dependence on SFR. The centre panel shows 
a weak and not significant correlation 
between SFR and the raw gradient slope 
on the mass-matched subsamples 
($r = +0.060$, $p = 0.276$). The right 
panel shows that the high-SFR group has 
a more metal-poor centre and a more 
positive residual gradient than the 
low-SFR group. The trend is in the same 
sense as for concentration, but the 
per-galaxy correlation is weak.}
\label{fig:sfr}
\end{figure*}

\begin{figure*}
\centering
\includegraphics[width=\hsize]{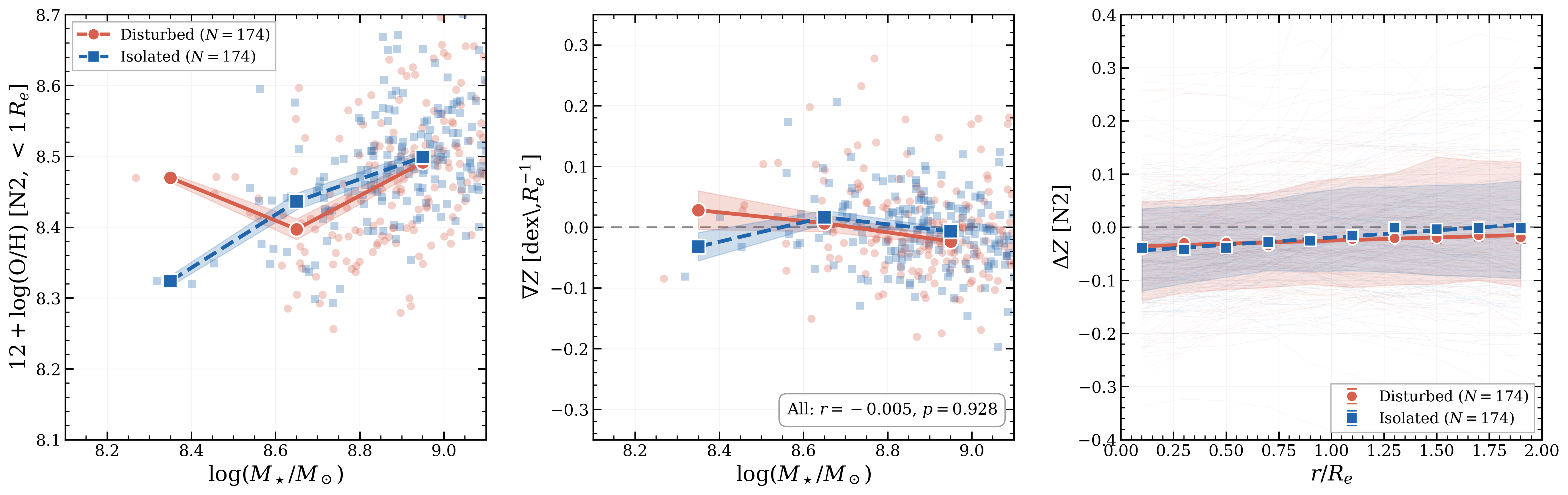}
\caption{As Figure~\ref{fig:conc}, but 
split by visual environmental 
classification into disturbed (red 
circles, solid) and isolated (blue 
squares, dashed) mass-matched subsamples. 
No significant difference appears in any 
panel: the left panel shows the same 
overall metallicity level for both groups, 
the centre panel shows no correlation 
between classification and the raw 
gradient slope ($r = -0.005$, 
$p = 0.928$), and the right panel shows 
residual profiles that overlap at all 
radii. This is a robust null result across 
all environment indicators tested, 
indicating that large scale environment 
does not measurably influence the radial 
distribution of metals.}
\label{fig:env}
\end{figure*}

We present four of these properties, 
light concentration, gas velocity 
dispersion, star formation rate, and 
environment, in the three-panel figures 
(Figures~\ref{fig:conc}--\ref{fig:env}). 
The three panels are the same in each 
figure. We split the sample at the median 
of the property into a higher and a lower 
group, shown throughout in contrasting 
colours. For light concentration the 
measurement is available for only 230 
galaxies, so its panels use two mass bins 
rather than the three used for the other 
properties. The \textit{left} panel shows 
the inner metallicity $Z(1\,R_{\rm e})$ 
against stellar mass, where 
$Z(1\,R_{\rm e})$ is the inverse variance 
weighted mean N2 metallicity over all bins 
within $1\,R_{\rm e}$. This panel recovers 
the mass-metallicity relation and shows 
whether the overall metallicity level 
depends on the property. The 
\textit{middle} panel shows the raw 
gradient slope $\nabla Z$ against stellar 
mass. A positive value means metallicity 
rises outward and a negative value means 
it falls outward, so this panel shows 
whether the property makes the gradient 
flatter, steeper, or inverted. The 
\textit{right} panel shows the median 
residual profiles $\Delta Z(r)$ for the 
two groups. A separation between the two 
groups here means the property shapes the 
radial metallicity structure independently 
of mass.

The correlation analysis uses a clean 
sample of 366 galaxies. We obtain it from 
the full sample of 382 by removing three 
galaxies at $z < 0.002$, for which 
$\Delta Z$ cannot be computed, and 
thirteen with poorly constrained gradient 
slopes 
($\sigma_{\nabla\Delta Z} > 0.05$\,dex\,$R_{\rm 
e}^{-1}$), which are noisy or non-linear 
profiles including several confirmed 
mergers. Concentration is the only 
property not measured for essentially the 
full sample, being available for 230 
galaxies. For the middle panels we run a 
complementary test on the raw slope 
$\nabla Z$ using mass-matched subsamples. 
We match following \citet{Wisz2025}, 
randomly discarding galaxies from the 
larger group in each mass bin until the 
two groups have an identical mass 
distribution. Any remaining difference 
between them then reflects the property 
rather than a residual dependence on mass. 
Table~\ref{tab:samples} collects the 
correlation coefficients for each 
property.

\begin{table*}
\centering
\caption{Correlation coefficients 
between each physical property and the 
residual gradient slope 
$\nabla\Delta Z$ over the clean sample 
of 366 galaxies, except for 
concentration which uses the 219 
galaxies with both a valid concentration 
measurement and a well-constrained 
gradient slope. The Pearson 
correlation $r_P$ and Spearman 
correlation $r_S$ are given alongside 
the coefficient of determination $r^2$ 
and the corresponding $p$-values 
$p_P$ and $p_S$. The $r(\nabla Z)$ 
column gives the complementary 
raw-gradient Pearson correlation on 
mass-matched subsamples. Stellar mass 
is not included in the $r(\nabla Z)$ 
column because mass-matching is not 
applicable when splitting by stellar 
mass itself.}
\label{tab:samples}
\begin{tabular}{lccccccc}
\toprule
Property & 
$r_P$ & $r_S$ & $r^2$ & 
$p_P$ & $p_S$ & $r(\nabla Z)$ \\
\midrule
Concentration & 
$+0.304$ & $+0.241$ & $0.092$ & 
$<0.001$ & $<0.001$ & $+0.217$ \\
Gas vel.\ disp. & 
$+0.204$ & $+0.289$ & $0.041$ & 
$<0.001$ & $<0.001$ & $+0.184$ \\
SFR & 
$+0.106$ & $+0.138$ & $0.011$ & 
$0.042$ & $0.008$ & $+0.060$ \\
Environment & 
$+0.006$ & $-0.026$ & $0.000$ & 
$0.914$ & $0.617$ & $-0.005$ \\
Stellar mass & 
$-0.078$ & $-0.116$ & $0.006$ & 
$0.134$ & $0.027$ & \dots \\
\bottomrule
\end{tabular}
\end{table*}

Light concentration $C$ shows the 
strongest and most significant correlation 
with the residual gradient 
(Pearson $r = 0.304$, Spearman $r = 0.241$, 
$r^2 = 0.092$, $p < 0.001$; 
Figure~\ref{fig:conc}), though it accounts 
for only 9 per cent of the observed 
variance in $\nabla\Delta Z$. Monte Carlo 
error propagation confirms that this 
result is robust to measurement 
uncertainties in the gradient slopes, 
with a 68 per cent confidence interval of 
$r \in [0.283, 0.313]$ 
(Appendix~\ref{app:mc}). The 
high-concentration galaxies have a more 
positive residual gradient than the 
low-concentration galaxies, with median 
$\nabla \Delta Z = +0.043$ against 
$+0.013$\,dex\,$R_{\rm e}^{-1}$. Their 
residual profiles show the clearest 
separation of any property we test, 
with the high-concentration group rising 
from $\Delta Z = -0.064$ at 
$0.1\,R_{\rm e}$ to $-0.026$ in the 
outskirts, while the low-concentration 
group rises from $-0.027$ to $+0.014$, 
suggesting that the centres of 
concentrated galaxies tend to be metal 
poor relative to what their high local 
stellar density predicts. The 
concentration result persists when a 
stricter resolution cut is applied 
($R_{\rm e} \geq 2\,\theta_{\rm PSF}$, 
$r = 0.234$, $p = 0.006$; 
Appendix~\ref{app:psf}), though the 
effect size is reduced.

Gas velocity dispersion $\sigma_{\rm gas}$ 
shows a statistically significant 
correlation with the residual gradient 
(Pearson $r = 0.204$, Spearman $r = 0.289$, 
$r^2 = 0.041$, $p < 0.001$; 
Figure~\ref{fig:sigmagas}), with 
high-dispersion galaxies tending to have 
more positive residual gradients. We note 
that $\sigma_{\rm gas}$ as measured from 
the MaNGA data shows some dependence on 
spatial resolution, suggesting that beam 
smearing may contribute to the measured 
dispersion in poorly resolved galaxies. 
The correlation with $\nabla\Delta Z$ 
should therefore be interpreted with some 
caution, and we return to this in 
Section~\ref{sec:disc_sigmagas}.

Star formation rate shows a marginal but 
significant correlation in both Pearson 
and Spearman tests 
(Pearson $r = 0.106$, $p = 0.042$; 
Spearman $r = 0.138$, $p = 0.008$; 
$r^2 = 0.011$; Figure~\ref{fig:sfr}), 
most likely through its link to stellar 
mass (Section~\ref{sec:disc_sfr}). 
Star formation rate accounts for only 
1 per cent of the variance in 
$\nabla\Delta Z$, and we regard this 
as a weak and tentative association. 
Total stellar mass shows no significant 
Pearson correlation with the residual 
gradient ($r_P = -0.078$, $p = 0.134$), 
though the Spearman test gives a 
marginally significant result 
($r_S = -0.116$, $p = 0.027$), 
suggesting a weak rank correlation 
that is not well described by a linear 
model.

Large scale environment has no detectable 
effect. Disturbed and isolated galaxies 
(Figure~\ref{fig:env}) differ neither in 
gradient slope ($r = 0.006$, $p = 0.914$) 
nor in their residual profiles, which 
overlap at all radii. This is a robust 
null result across both Pearson and 
Spearman tests and across all environment 
indicators from the GEMA catalogue, 
indicating that large scale environment 
does not measurably influence the radial 
distribution of metals in our sample.

Figure~\ref{fig:correlations} summarises 
the full set of coefficients, and we 
discuss the physical interpretation in 
Section~\ref{sec:discussion}.

\section{Discussion}
\label{sec:discussion}

\subsection{Flat Metallicity Gradients in 
Low-Mass Galaxies}
\label{sec:disc_flat}
We find that the metallicity gradients 
across our low-mass MaNGA galaxies are 
flat on average 
(Figure~\ref{fig:gradients}, 
Section~\ref{sec:gradients}). The median 
slope is $-0.007$\,dex\,$R_{\rm e}^{-1}$, 
with a galaxy-to-galaxy scatter of 
$0.071$\,dex\,$R_{\rm e}^{-1}$. We note 
that beam smearing in the less well 
resolved galaxies may contribute to this 
flattening; when a stricter resolution 
cut of $R_{\rm e} \geq 2\,\theta_{\rm PSF}$ 
is applied the median slope shifts to 
$-0.026$\,dex\,$R_{\rm e}^{-1}$ 
(Appendix~\ref{app:psf}), which may be 
a closer approximation to the intrinsic 
population median. In either case the 
slopes are distributed broadly about zero, 
consistent with a balance between mild 
negative and mild positive gradients 
rather than uniformly flat profiles.

This behaviour differs from that of 
massive spiral galaxies, in which 
inside-out disk growth produces clearly 
negative gradients 
\citep{Sanchez2014, Belfiore2017}. Our 
low-mass galaxies instead show flat and 
broadly scattered gradients, with no 
strong preference for a negative slope. 
Our results extend the gradient flattening 
that \citet{Belfiore2017} saw toward the 
low-mass end of MaNGA to a homogeneous 
sample that reaches well into the dwarf 
galaxy regime, consistent with recent 
dedicated studies of the dwarf regime 
\citep{Li2025, Vale2025}.

A small number of galaxies stand out with 
strongly positive, inverted gradients. 
Roughly 11 of our 382 galaxies (3 per 
cent) have $\nabla Z > 0.15$\,dex\,$R_{\rm 
e}^{-1}$, and the most extreme reaches 
$+0.28$\,dex\,$R_{\rm e}^{-1}$. Positive 
gradients of this magnitude have been 
reported in other low-mass samples and 
attributed to the inflow of metal-poor 
gas to the central regions 
\citep{Grossi2020}, though we do not find 
a statistically significant association 
between positive gradients and disturbed 
morphology in our sample 
(Section~\ref{sec:correlations}). 
Alternative explanations, including 
residual beam-smearing effects or 
stochastic star formation histories, 
cannot be excluded.

\subsection{Concentration as a Driver of 
Metallicity Gradients}
\label{sec:disc_concentration}
Light concentration $C$ shows the most 
significant association with the 
metallicity gradient, both in the raw 
gradient slope $\nabla Z$ (middle panel 
of Figure~\ref{fig:conc}) and in the 
stellar-mass-subtracted residual slope 
$\nabla \Delta Z$ (right panel). We split 
the sample at the median $C = 2.909$. The 
left panel shows that the high- and 
low-$C$ groups follow the same 
mass-metallicity relation, so the overall 
metallicity level does not vary with 
concentration. The middle panel shows 
that the high-$C$ group tends to have 
more positive raw slopes ($r = 0.217$, 
$p = 0.002$). The right panel shows the 
same trend in the residual 
(Pearson $r = 0.304$, Spearman $r = 0.241$, 
$r^2 = 0.092$, $p < 0.001$), though 
concentration accounts for only 9 per 
cent of the variance in 
$\nabla \Delta Z$. Monte Carlo error 
propagation gives a 68 per cent 
confidence interval of 
$r \in [0.283, 0.313]$ 
(Appendix~\ref{app:mc}), confirming the 
result is stable to measurement 
uncertainties. The high-$C$ group has 
a median residual slope of 
$\nabla \Delta Z = +0.043$ against 
$+0.013$\,dex\,$R_{\rm e}^{-1}$ for the 
low-$C$ group. The raw and residual 
correlations agree, suggesting the trend 
is not an artefact of the mass 
subtraction.

We note that the two median residual 
profiles in the right panel of 
Figure~\ref{fig:conc} differ primarily 
in normalisation rather than slope, with 
the high-$C$ group rising from 
$\Delta Z = -0.064$ at $0.1\,R_{\rm e}$ 
to $-0.026$ in the outskirts and the 
low-$C$ group rising from $-0.027$ to 
$+0.014$. This is expected: stacking 
many galaxies averages out the individual 
gradient differences, so the separation 
between the group median profiles 
reflects the difference in median 
$\Delta Z$ level rather than the 
difference in median slope. The 
per-galaxy slope distribution, which is 
the quantity entering the correlation 
analysis, captures this slope difference 
directly.

The result persists when a stricter PSF 
resolution cut is applied, though with a 
reduced effect size 
($R_{\rm e} \geq 2\,\theta_{\rm PSF}$, 
$r = 0.234$, $p = 0.006$; 
Appendix~\ref{app:psf}).

A positive residual gradient means the 
centre is metal poor relative to what its 
high local stellar mass surface density 
predicts. In concentrated galaxies the 
central stellar density is high, so the 
$Z$--$\Sigma_{\star}$ relation predicts a 
high central metallicity, yet the measured 
central gas-phase metallicity falls below 
it. The stars in the centre are metal 
rich, as their high density implies, while 
the gas is relatively metal poor, 
suggesting that the central gas may have 
been diluted by an inflow of 
lower-metallicity material. One possible 
explanation is the inflow of metal-poor 
gas to the centres of concentrated 
systems, which could dilute the central 
gas-phase abundance while the stellar 
component that sets $\Sigma_{\star}$ 
remains metal rich. This is broadly 
consistent with \citet{CanoD2025}, who 
find that stellar population gradients in 
MaNDala galaxies reflect the degree of 
inside-out versus outside-in formation, 
with more concentrated systems showing 
more centrally concentrated star formation 
histories. We caution, however, that the modest 
effect size means that this interpretation 
should be treated as tentative pending 
confirmation with higher-resolution data. 
We show in Appendix~\ref{app:psf} that 
the concentration correlation is not a 
spatial resolution artefact, with the 
partial correlation at fixed 
$R_{\rm e}/\theta_{\rm PSF}$ actually 
strengthening to $r = 0.359$.

\subsection{A Connection with Gas Velocity 
Dispersion}
\label{sec:disc_sigmagas}
The gas velocity dispersion within 
$1\,R_{\rm e}$, $\sigma_{\rm gas}$ 
(Figure~\ref{fig:sigmagas}), shows a 
statistically significant association 
with the residual gradient when the 
sample is split at the median of 
$72.5$\,km\,s$^{-1}$. The left panel 
shows that the high- and 
low-$\sigma_{\rm gas}$ groups follow the 
same mass-metallicity relation, so the 
overall metallicity level does not vary 
with $\sigma_{\rm gas}$. The middle panel 
shows that the high-dispersion group 
tends to have more positive raw gradient 
slopes ($r = 0.184$, $p = 0.001$), with 
little variation with stellar mass. The 
right panel shows the same trend in the 
residual (Pearson $r = 0.204$, 
Spearman $r = 0.289$, $r^2 = 0.041$, 
$p < 0.001$). The high-dispersion group 
has inner $\Delta Z = -0.048$ rising to 
$+0.014$ in the outskirts, with median 
$\nabla \Delta Z = +0.045$ against 
$+0.004$\,dex\,$R_{\rm e}^{-1}$ for the 
low-dispersion group.

We note that $\sigma_{\rm gas}$ as 
measured from the MaNGA data shows some 
dependence on spatial resolution, and 
beam smearing may contribute to the 
measured dispersion in poorly resolved 
galaxies. This limits our ability to draw 
strong physical conclusions from this 
correlation, and the result should be 
regarded as suggestive rather than 
definitive pending higher-resolution 
observations.

If the correlation is physical, one 
possible interpretation is that high 
$\sigma_{\rm gas}$ traces turbulence 
driven by strong stellar feedback, which 
drives outflows that transport metals 
from the centre toward the outskirts, 
raising the outer metallicity relative 
to the centre and producing the more 
positive residual gradient. Concentration 
and $\sigma_{\rm gas}$ are uncorrelated 
($r = 0.028$, $p = 0.677$), so if both 
effects are physical they act as 
independent channels rather than a single 
underlying driver.

\subsection{The Absence of Environmental 
Effects}
\label{sec:disc_environment}
We find no significant association between 
large scale environment and the residual 
gradient slope (Figure~\ref{fig:env}). 
The number of neighbours, the group size, 
the group membership, and the visual 
classification into isolated and disturbed 
galaxies all fail to produce a significant 
correlation in either the Pearson or 
Spearman tests 
(Figure~\ref{fig:correlations}). Our 
groups are well populated on both sides 
(189 disturbed and 177 isolated  in the clean sample), 
so this is a robustly constrained null 
rather than a limitation of sample size. 
We note that this result may depend on 
how environment is defined. A quantitative 
classification based on the CAS 
parameters, for example, may not give the 
same result as our visual morphology 
classification.

Across the mass range we probe, these 
galaxies are predominantly low-density 
field systems. For such galaxies the 
internal baryonic physics, which includes 
the spatial concentration of star 
formation and the turbulent state of the 
gas, may dominate over external processes 
in shaping the radial distribution of 
metals. Ram-pressure stripping and tidal 
interactions \citep{Boselli2022} may 
become important at higher densities or 
in cluster environments, but our sample 
contains too few cluster members to test 
this regime directly.

\subsection{The Star Formation Rate 
Connection}
\label{sec:disc_sfr}
The total star formation rate, split at 
the sample median (Figure~\ref{fig:sfr}), 
shows only a marginal association with 
the residual gradient slope. The left 
panel shows that the high- and low-SFR 
groups follow the same mass-metallicity 
relation, so the overall metallicity level 
does not vary with SFR. The middle panel 
shows that the high-SFR group has 
slightly more positive raw gradient slopes 
than the low-SFR group ($r = 0.060$, 
$p = 0.276$). The right panel shows the 
residual profiles. The high-SFR group 
has a more metal-poor centre, with inner 
$\Delta Z = -0.062$ rising to $-0.024$ 
in the outskirts, and a median residual 
slope of 
$\nabla \Delta Z = +0.036$\,dex\,$R_{\rm 
e}^{-1}$. The low-SFR group has a less 
depressed centre (inner 
$\Delta Z = -0.018$) and a shallower 
median slope of 
$+0.010$\,dex\,$R_{\rm e}^{-1}$. This 
is qualitatively similar to the trend 
seen for gas velocity dispersion, and 
could be consistent with a picture in 
which higher SFR drives stronger feedback 
and outflows that transport metals 
outward.

We treat this result with caution. The 
per-galaxy correlation is weak 
(Pearson $r = 0.106$, $p = 0.042$; 
Spearman $r = 0.138$, $p = 0.008$; 
$r^2 = 0.011$), with SFR accounting for 
only 1 per cent of the variance in 
$\nabla\Delta Z$. The trend is also 
difficult to disentangle from stellar 
mass, since SFR and stellar mass are 
correlated through the star-forming main 
sequence and a weak residual mass effect 
could appear as a weak SFR correlation. 
We therefore do not claim an independent 
physical effect of star formation rate on 
metallicity gradients, and regard the 
correlation as suggestive at most.

\subsection{Other Parameters}
\label{sec:disc_other}
\begin{figure*}
\centering
\includegraphics[width=\hsize]{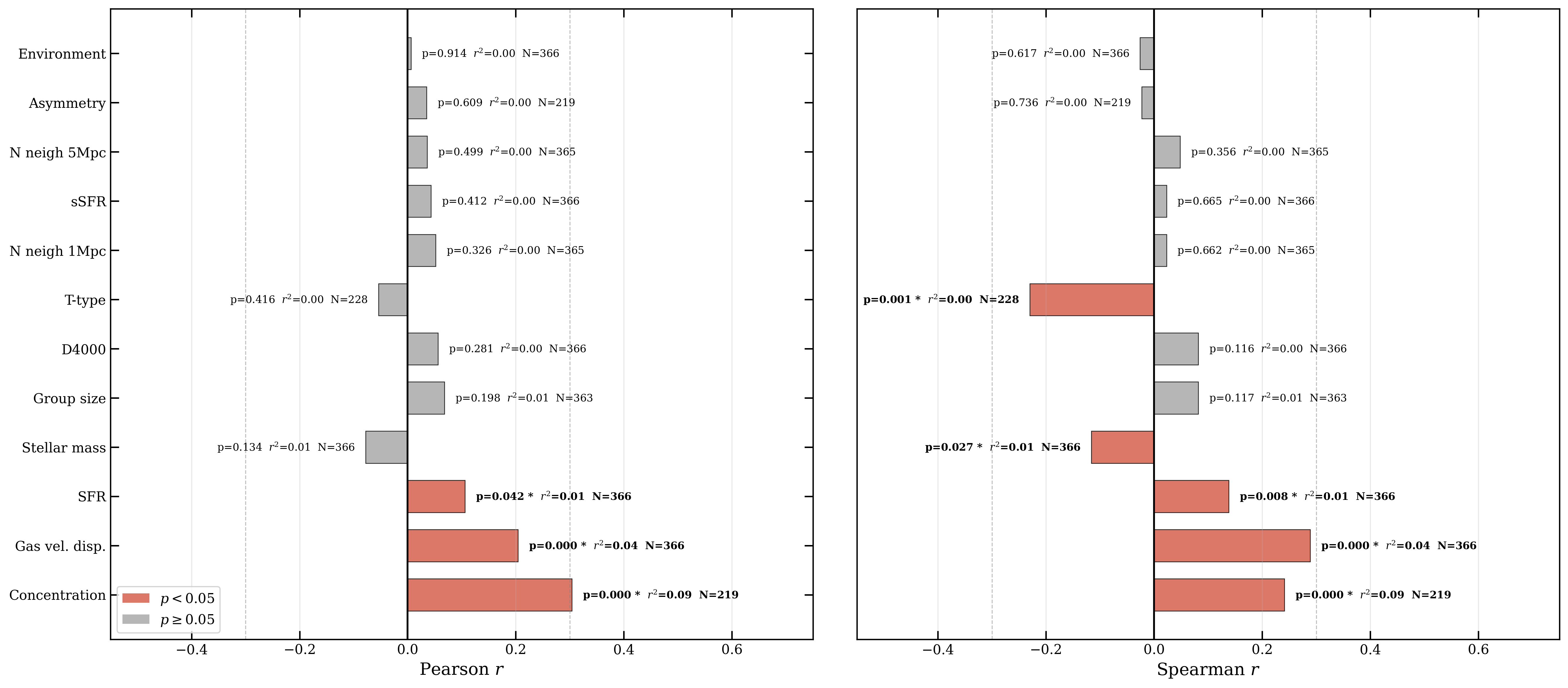}
\caption{Summary of the Pearson (left) 
and Spearman (right) correlation 
coefficients between the residual 
gradient slope $\nabla \Delta Z$ 
(Section~\ref{sec:gradient}) and all the 
physical properties we test, over the 
clean sample of 366 galaxies, sorted by 
absolute Pearson $r$. Coloured bars are 
significant at $p < 0.05$ and grey bars 
are not. The $p$-value, coefficient of 
determination $r^2$, and sample size $N$ 
are annotated for each property. Light 
concentration shows the strongest and 
most significant association in both 
tests. Gas velocity dispersion shows a 
significant correlation that should be 
interpreted with caution given its 
sensitivity to spatial resolution 
(Section~\ref{sec:disc_sigmagas}). 
Morphological T-type shows a significant 
Spearman correlation but not a 
significant Pearson correlation, 
suggesting a weak non-linear association 
connected to the concentration result 
(Section~\ref{sec:disc_other}). Large 
scale environment and the remaining 
properties show no significant effect in 
either test. These are the headline 
correlations with the residual slope, 
distinct from the raw-slope correlations 
annotated in the middle panels of the 
three-panel figures.}
\label{fig:correlations}
\end{figure*}

Figure~\ref{fig:correlations} shows the 
Pearson and Spearman correlation 
coefficients and $r^2$ effect sizes for 
every property we tested, with the 
significant ones marked in colour. 
Concentration shows the strongest and 
most significant association, star 
formation rate sits at the margin of 
significance, and gas velocity dispersion 
shows a significant correlation that 
should be interpreted with caution given 
its sensitivity to spatial resolution 
(Section~\ref{sec:disc_sigmagas}). All 
other parameters show no significant 
effect in both Pearson and Spearman tests, 
with one exception noted below.

The environmental measures from the GEMA 
catalogue, which are the group size, the 
number of neighbours, and the group 
membership 
(Section~\ref{sec:ancillary}), are all 
consistent with no correlation. This 
agrees with the absence of environmental 
effects in 
Section~\ref{sec:disc_environment}. The 
internal properties, which are the 
specific star formation rate, the D4000 
break strength, together with asymmetry 
(Section~\ref{sec:visual_class}), are 
also uncorrelated with the residual 
gradient in both tests.

The morphological T-type shows no 
significant Pearson correlation 
($r = -0.054$, $p = 0.416$) but a 
significant Spearman correlation 
($r_S = -0.230$, $p < 0.001$, $N = 228$), 
indicating a weak rank correlation between 
earlier morphological type and more positive 
residual gradients. This result is likely 
connected to the concentration result, 
since concentration and T-type are 
physically linked, more concentrated 
galaxies tend to have earlier morphological 
types. We therefore do not treat this as 
an independent result but note it as 
consistent with the concentration finding.

Taken together, these results suggest 
that the radial distribution of metals 
in low-mass galaxies is more closely 
linked to internal structural properties 
than to morphology, star formation 
history, or large scale environment, 
though the modest effect sizes indicate 
that no single parameter is a dominant 
driver.

\subsection{Comparison with Previous Work}
\label{sec:disc_literature}
Our results broadly agree with the 
stellar population gradient results of 
\citet{CanoD2025} for the overlapping 
MaNDala sample. They find mildly negative 
luminosity-weighted gradients and flat to 
slightly positive mass-weighted stellar 
metallicity gradients, which are 
qualitatively consistent with our flat 
gas-phase gradients.

The average gradient is flat, but when 
we split the sample by properties that 
may trace internal gas flows and stellar 
feedback, namely concentration, gas 
velocity dispersion, and star formation 
rate, tentative trends appear. Galaxies 
with higher concentration, higher gas 
velocity dispersion, and higher star 
formation rate tend to show more positive 
residual gradients, though the effect 
sizes are modest ($r^2 \leq 0.092$) and 
no single parameter accounts for more 
than 9 per cent of the variance in 
$\nabla\Delta Z$. We note that the gas 
velocity dispersion result should be 
treated with caution given its 
sensitivity to spatial resolution 
(Section~\ref{sec:disc_sigmagas}). One 
possible interpretation is a picture in 
which feedback-driven outflows and 
inflows continually redistribute metals 
in these shallow potential wells 
\citep{Mercado2021}, with galaxies caught 
at different phases of this cycle showing 
different gradient slopes. The flat 
average would then be the long-term 
outcome, while the property trends pick 
out galaxies in the act of redistributing 
their metals. We caution, however, that 
the modest effect sizes and the 
sensitivity of some results to the PSF 
resolution cut mean that this 
interpretation remains speculative, and 
confirmation with larger samples and 
higher-resolution data would be valuable. 
Our results are also broadly consistent 
with recent dedicated IFU studies of 
dwarf galaxies \citep{Li2025, Vale2025}, 
which find similarly flat or mildly 
negative gradients in this mass regime.

\section{Conclusions}
\label{sec:conclusions}
We have presented spatially resolved 
gas-phase metallicity gradients for 382 
star-forming low-mass galaxies selected 
from the complete MaNGA DR17 survey 
\citep{Abdurrouf2022}. This is the 
largest spatially resolved spectroscopic 
study of dwarf galaxy metallicity 
gradients to date. Our main conclusions 
are as follows.

\begin{enumerate}

\item Metallicity gradients are flat on 
average, with a median slope of 
$-0.007$\,dex\,$R_{\rm e}^{-1}$ and a 
galaxy-to-galaxy scatter of 
$0.071$\,dex\,$R_{\rm e}^{-1}$. The 
slopes are distributed broadly about 
zero, consistent with a balance of mild 
positive and negative gradients rather 
than uniformly flat profiles. Beam 
smearing in the less well resolved 
galaxies may contribute to this 
flattening; a stricter resolution cut 
($R_{\rm e} \geq 2\,\theta_{\rm PSF}$, 
$N = 254$) yields a more negative median 
slope of $-0.026$\,dex\,$R_{\rm e}^{-1}$ 
(Appendix~\ref{app:psf}).

\item The N2 calibration of 
\citet{PettiniPagel2004} recovers a 
significant mass-metallicity relation 
for our sample (Pearson $r = 0.444$, 
$r^2 = 0.197$, $N = 380$). For the 
subset of 379 galaxies with a valid 
O3N2 measurement, one galaxy 
(8934-9102) is excluded because its 
N2-derived metallicity is unphysically 
high ($12 + \log(\mathrm{O/H}) = 9.20$), 
likely reflecting a failed fit rather 
than a true abundance. On this subsample 
the N2 correlation strengthens to 
$r = 0.473$, while O3N2 gives 
$r = 0.380$, confirming N2 as the more 
reliable indicator for this low-mass 
regime (Appendix~\ref{app:calibration}).

\item We find a significant positive 
correlation between gas-phase metallicity 
and local stellar mass surface density 
$\Sigma_{\star}$ (Pearson $r = 0.371$, 
$r^2 = 0.138$), indicating that denser 
regions tend to be more metal rich, 
though local surface density accounts 
for only 14 per cent of the observed 
variance. The linearity of this relation 
and a small systematic correction applied 
to the residual profiles are discussed in 
Appendix~\ref{app:nonlinear}.

\item After removing the $\Sigma_{\star}$ 
dependence, the residual metallicity 
$\Delta Z$ retains a clear mass offset 
at all radii. More massive galaxies tend 
to be more metal rich than their local 
density predicts, and less massive 
galaxies more metal poor. Total stellar 
mass therefore sets the overall 
enrichment level, but shows no 
significant Pearson correlation with the 
residual gradient slope ($r_P = -0.078$, 
$r^2 = 0.006$, $p = 0.134$), though the 
Spearman test gives a marginally 
significant result ($r_S = -0.116$, 
$p = 0.027$).

\item The strongest association with the 
residual gradient slope $\nabla \Delta Z$ 
is light concentration $C$ (Pearson 
$r = 0.304$, Spearman $r = 0.241$, 
$r^2 = 0.092$, $p < 0.001$), though 
concentration accounts for only 9 per 
cent of the variance. More concentrated 
galaxies tend to have more positive 
residual gradients, with centres that 
are metal poor relative to what their 
high local stellar density predicts. 
Monte Carlo error propagation confirms 
the result is robust to measurement 
uncertainties (68 per cent CI: 
$r \in [0.283, 0.313]$; 
Appendix~\ref{app:mc}). The result 
persists with a stricter PSF resolution 
cut, though with reduced effect size 
($r = 0.234$, $p = 0.006$; 
Appendix~\ref{app:psf}).

\item Gas velocity dispersion 
$\sigma_{\rm gas}$ shows a statistically 
significant positive correlation 
(Pearson $r = 0.204$, Spearman $r = 0.289$, 
$r^2 = 0.041$, $p < 0.001$), and star 
formation rate shows a weaker association 
(Pearson $r = 0.106$, Spearman $r = 0.138$, 
$r^2 = 0.011$, $p = 0.042$). We caution 
that the $\sigma_{\rm gas}$ result shows 
some sensitivity to spatial resolution 
and should be interpreted cautiously 
pending higher-resolution observations 
(Section~\ref{sec:disc_sigmagas}).

\item Large scale environment has no 
significant effect on the residual 
gradient slope. The visual classification 
into disturbed and isolated galaxies 
(189 and 177 galaxies respectively), 
and all quantitative environment 
indicators from the GEMA catalogue, show 
no significant correlation in either 
Pearson or Spearman tests, which is a 
robust null result.

\end{enumerate}

Taken together, these results suggest 
that internal structural properties, 
particularly light concentration, are 
more closely linked to the radial 
distribution of metals in low-mass 
galaxies than stellar mass, star 
formation history, or large scale 
environment. The modest effect sizes 
($r^2 \leq 0.092$) indicate that no 
single parameter is a dominant driver 
and that much of the galaxy-to-galaxy 
scatter remains unexplained. The flat 
average gradient combined with the 
tentative trends in concentration, gas 
velocity dispersion, and star formation 
rate is broadly consistent 
with a picture in which feedback-driven 
processes redistribute metals in shallow 
potential wells \citep{Mercado2021}, 
though confirmation with larger samples 
and higher spatial resolution would be 
needed to establish this interpretation 
on firmer ground. Large scale environment 
plays no detectable role across the mass 
range we probe, suggesting that these 
galaxies are shaped primarily by their 
own internal evolution rather than by 
their surroundings.

\begin{acknowledgements}
We thank the anonymous referee for a 
thorough and constructive report that 
significantly improved the quality and 
rigour of this work.

A.D. acknowledges support from ANID through 
the Beca de Doctorado Nacional, Folio 
21260921. S.S. and R.C. thank ANRF for financial 
support through the SERB-SURE grants 
(SUR/2022/001789 and SUR/2022/001503, respectively), and IUCAA 
for their hospitality and usage of their 
facilities during their stay at different 
times as part of the university 
associateship program. R.C. thanks ANRF 
for an ARG grant 
(ANRF/ARG/2025/003315/PS).

This work uses data from the Sloan Digital 
Sky Survey IV, funded by the Alfred P. 
Sloan Foundation, the U.S. Department of 
Energy Office of Science, and the 
Participating Institutions 
(\url{https://www.sdss4.org}). It also 
uses data from the DESI Legacy Imaging 
Surveys (DECaLS, BASS, and MzLS) and 
NEOWISE.

This research made use of 
\textsc{astropy} \citep{Astropy2018}, 
\textsc{numpy} \citep{Numpy2020}, 
\textsc{scipy} \citep{Scipy2020}, and 
\textsc{matplotlib} \citep{Matplotlib2007}.
\end{acknowledgements}

\bibliographystyle{aa}
\bibliography{references}
\begin{appendix}

\section{PSF Smearing Analysis}
\label{app:psf}

To verify that our gradient measurements are 
not significantly affected by beam smearing 
from the MaNGA point spread function 
(PSF, $\theta_{\rm PSF} = 2.5$\,arcsec FWHM), 
we carried out six tests.

\textit{Unresolved versus resolved galaxies.}
We compared the distribution of gradient slopes 
for spatially unresolved galaxies 
($R_{\rm e} < 1 \times \theta_{\rm PSF}$, 
$N = 8$) with that of the resolved galaxies 
($R_{\rm e} \geq 1 \times \theta_{\rm PSF}$, 
$N = 382$). Unresolved galaxies show a 
systematic bias toward positive gradient slopes 
(median $+0.034$\,dex\,$R_{\rm e}^{-1}$) 
relative to resolved galaxies 
(median $-0.007$\,dex\,$R_{\rm e}^{-1}$; 
Mann--Whitney $p = 0.015$), confirming that 
the $R_{\rm e} \geq 1 \times \theta_{\rm PSF}$ 
cut is necessary. This is the expected 
signature of beam smearing, where light from 
the metal-rich central regions is spread into 
outer bins by the PSF, artificially producing 
a spuriously positive gradient 
\citep{Acharyya2020}.

\begin{figure*}
\centering
\includegraphics[width=\hsize]{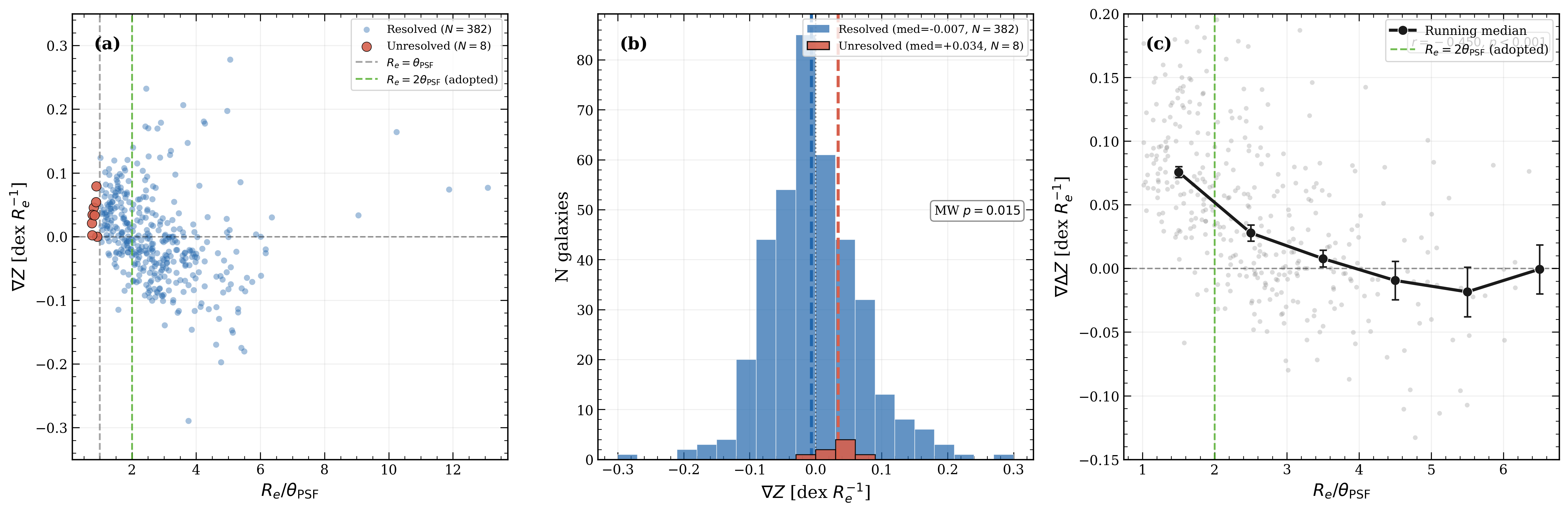}
\caption{PSF beam smearing analysis. 
\textit{Panel (a):} metallicity gradient 
slope $\nabla Z$ as a function of 
$R_{\rm e}/\theta_{\rm PSF}$ for resolved 
(blue, $N = 382$) and unresolved (red, 
$N = 8$) galaxies. The dashed vertical 
lines mark $R_{\rm e} = 1\,\theta_{\rm PSF}$ 
(grey) and $R_{\rm e} = 2\,\theta_{\rm PSF}$ 
(green, adopted robustness cut). 
\textit{Panel (b):} histogram of $\nabla Z$ 
for unresolved (red) and resolved (blue) 
galaxies. Unresolved galaxies are biased 
toward positive slopes 
(median $+0.034$\,dex\,$R_{\rm e}^{-1}$) 
relative to resolved galaxies 
(median $-0.007$\,dex\,$R_{\rm e}^{-1}$; 
Mann--Whitney $p = 0.015$). 
\textit{Panel (c):} residual gradient slope 
$\nabla\Delta Z$ as a function of 
$R_{\rm e}/\theta_{\rm PSF}$ for the full 
resolved sample. The black line shows the 
running median. Even within the resolved 
sample the gradient correlates with 
resolution ($r = -0.450$, $p < 0.001$), 
motivating the stricter 
$R_{\rm e} \geq 2\,\theta_{\rm PSF}$ cut 
shown by the green dashed line.}
\label{fig:psf}
\end{figure*}

\textit{Residual dependence on resolution.}
Panel~(c) of Figure~\ref{fig:psf} shows the 
residual metallicity gradient slope 
$\nabla\Delta Z$ as a function of 
$R_{\rm e}/\theta_{\rm PSF}$ for the full 
resolved sample. Even within the resolved 
sample we find a significant negative 
correlation ($r = -0.450$, $p < 0.001$), 
meaning that galaxies closer to the PSF limit 
still have systematically more positive 
residual gradients. This residual dependence 
motivates a stricter resolution cut. The 
trend flattens noticeably above 
$R_{\rm e} \geq 2 \times \theta_{\rm PSF}$, 
which we adopt as a robustness threshold.

\textit{Stability of results with stricter 
resolution cuts.}
Table~\ref{tab:psf_stability} shows how our 
key results change as we apply progressively 
stricter resolution cuts. The concentration 
correlation remains significant at 
$R_{\rm e} \geq 2 \times \theta_{\rm PSF}$ 
($r = 0.234$, $r^2 = 0.055$, $p = 0.006$, 
$N = 254$) and the median gradient slope 
stabilises around $-0.026$ to 
$-0.034$\,dex\,$R_{\rm e}^{-1}$, more 
negative than the full-sample value of 
$-0.007$\,dex\,$R_{\rm e}^{-1}$ because 
beam smearing was artificially flattening 
the gradients in the less well resolved 
galaxies. We therefore adopt the full 382 
galaxy sample as our primary sample 
throughout the paper, and note that the 
main scientific conclusions are robust to 
applying the stricter 
$R_{\rm e} \geq 2 \times \theta_{\rm PSF}$ 
cut.

\textit{Fitting gradients over 
PSF-unaffected radial ranges.}
For each galaxy we identified the bins 
whose centres lie within one PSF FWHM 
of the galaxy centre and refitted the 
gradients using only the outer bins, 
excluding on average 2.2 bins per 
galaxy. If beam smearing were confined 
to the innermost bins, this test would 
eliminate the bias. The correlation 
between $\nabla\Delta Z$ and 
$R_{\rm e}/\theta_{\rm PSF}$ actually 
strengthens slightly with this approach 
($r = -0.479$ versus $r = -0.450$ for 
the full radial range), indicating that 
the PSF effect is not confined to the 
innermost bins but reflects a more 
general smearing of the gradient 
measurement in poorly resolved galaxies. 
This reinforces the need for the 
stricter resolution cut rather than 
resolving the issue through bin 
exclusion alone.

\textit{Inclination.}
We find no significant correlation between 
galaxy inclination and either $\nabla Z$ 
($r = -0.036$, $p = 0.49$) or 
$\nabla\Delta Z$ ($r = -0.009$, $p = 0.86$), 
confirming that inclination effects do not 
bias our gradient measurements.

\textit{Partial correlation of concentration 
at fixed resolution.}
A potential concern is that the concentration 
correlation with $\nabla\Delta Z$ could be 
driven by a hidden dependence on spatial 
resolution, if more concentrated galaxies 
happened to be better resolved. We tested 
this by first checking whether concentration 
$C$ correlates with $R_{\rm e}/\theta_{\rm 
PSF}$ in our sample. We find no significant 
correlation ($r = 0.038$, $p = 0.572$, 
$N = 219$), confirming that concentrated 
galaxies are not preferentially better or 
worse resolved than less concentrated ones. 
We then computed the partial correlation 
between $C$ and $\nabla\Delta Z$ after 
removing the linear dependence of both 
quantities on $R_{\rm e}/\theta_{\rm PSF}$. 
The partial correlation is 
$r = 0.359$ ($r^2 = 0.129$, $p < 0.001$), 
stronger than the raw correlation of 
$r = 0.304$, demonstrating that the 
concentration correlation is not a 
resolution artefact.

This finding may appear to be in tension with 
the monotonic decline of the concentration 
correlation in 
Table~\ref{tab:psf_stability} 
($r = 0.304 \to 0.294 \to 0.234 \to 
0.141 \to 0.123$ as the resolution cut 
tightens). However the two tests measure 
different things. The partial correlation 
removes the smooth linear trend of 
$\nabla\Delta Z$ with 
$R_{\rm e}/\theta_{\rm PSF}$ while 
retaining the full sample of $N = 219$ 
galaxies with concentration measurements. 
The stability table instead restricts to 
progressively smaller and better-resolved 
subsamples, reducing $N$ from 219 to 181 
to 137 to 103 to 66. The weakening in 
Table~\ref{tab:psf_stability} is therefore 
consistent with reduced statistical power 
at smaller sample sizes rather than a 
physical resolution dependence, and the 
partial correlation provides the more 
direct test of whether spatial resolution 
drives the result.

The results are shown in 
Figure~\ref{fig:psf}.

\begin{table*}[t]
\centering
\caption{Stability of key results with 
progressively stricter PSF resolution 
cuts. Columns give the resolution 
threshold, total number of galaxies, 
median metallicity gradient slope, and 
Pearson correlation coefficients with 
their effect sizes ($r^2$) and $p$-values 
for light concentration $C$, star 
formation rate, and visual environment 
classification (Mann--Whitney $p$-value).}
\label{tab:psf_stability}
\begin{tabular}{lcccccccc}
\hline\hline
Cut & $N$ & $\langle\nabla Z\rangle$ & 
$r_C$ & $r_C^2$ & $p_C$ & 
$r_{\rm SFR}$ & $p_{\rm SFR}$ & 
$p_{\rm env}$ \\
 & & (dex\,$R_e^{-1}$) & & & & & & \\
\hline
${\geq}1.0\,\theta_{\rm PSF}$ & 382 & 
$-0.007$ & $0.304$ & $0.092$ & $<0.001$ & 
$0.106$ & $0.042$ & $0.616$ \\
${\geq}1.5\,\theta_{\rm PSF}$ & 323 & 
$-0.020$ & $0.294$ & $0.087$ & $<0.001$ & 
$0.123$ & $0.031$ & $0.985$ \\
${\geq}2.0\,\theta_{\rm PSF}$ & 254 & 
$-0.026$ & $0.234$ & $0.055$ & $0.006$ & 
$0.149$ & $0.021$ & $0.331$ \\
${\geq}2.5\,\theta_{\rm PSF}$ & 194 & 
$-0.034$ & $0.141$ & $0.020$ & $0.154$ & 
$0.153$ & $0.040$ & $0.612$ \\
${\geq}3.0\,\theta_{\rm PSF}$ & 131 & 
$-0.031$ & $0.123$ & $0.015$ & $0.326$ & 
$0.058$ & $0.529$ & $0.473$ \\
\hline\hline
\end{tabular}
\end{table*}

%\FloatBarrier

\section[Linearity of the Z-Sigma 
relation]{Linearity of the 
$Z$--$\Sigma_\star$ Relation}
\label{app:nonlinear}
\begin{figure}
\centering
\includegraphics[width=\hsize]{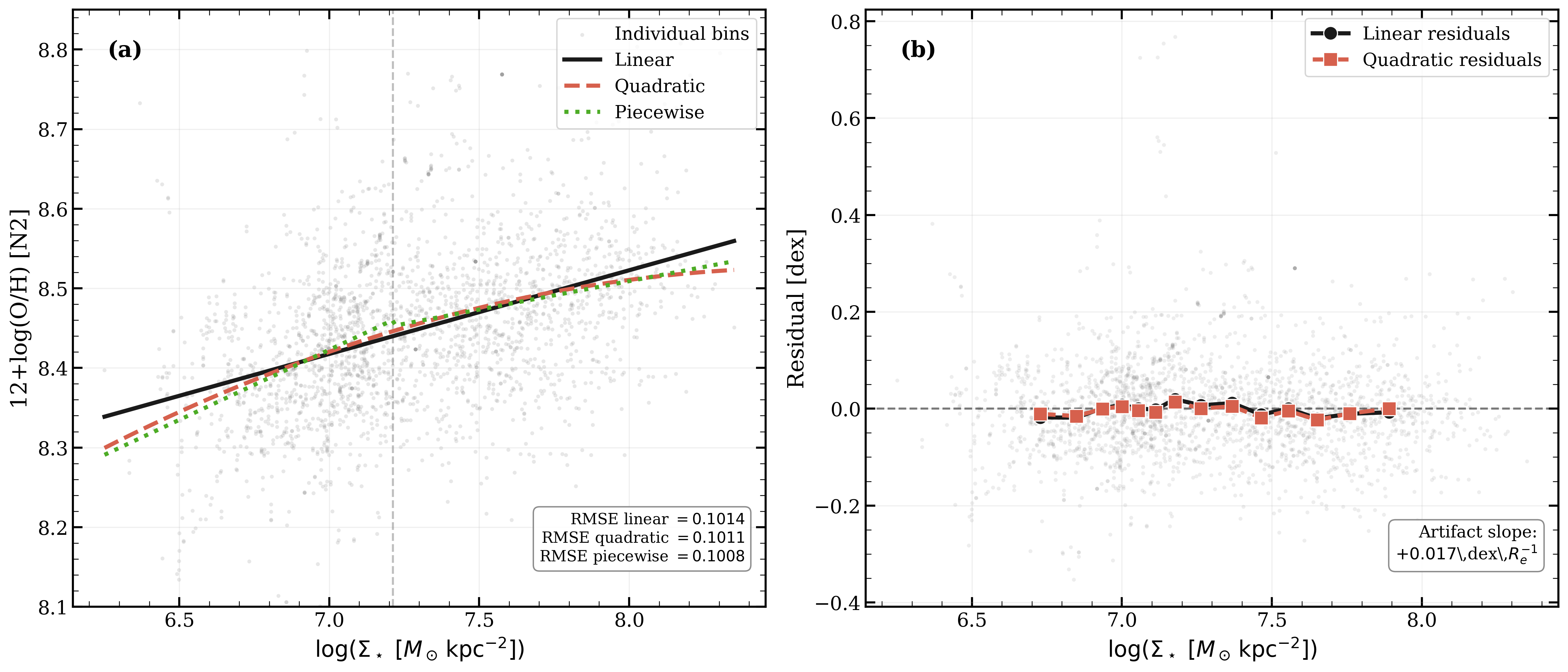}
\caption{Test of the linearity of the 
$Z$--$\log\Sigma_\star$ relation. 
\textit{Left:} comparison of linear 
(black), quadratic (red dashed), and 
piecewise linear (green dotted) fits to 
the 3608 valid bins. The three fits are 
visually indistinguishable. 
\textit{Right:} running median of the 
residuals from the linear (black) and 
quadratic (red) fits as a function of 
$\log\Sigma_\star$. The linear fit shows 
a small systematic trend of 
$+0.017$\,dex\,$R_e^{-1}$ which we 
correct for as described in the text.}
\label{fig:nonlinear}
\end{figure}

The residual metallicity $\Delta Z$ is 
defined by subtracting the prediction of 
a global linear fit to the 
$Z$--$\log\Sigma_\star$ relation 
(Section~\ref{sec:residual}). The results 
of the linearity test are shown in 
Figure~\ref{fig:nonlinear}.

First, we compared the root-mean-square 
residuals (RMSE) of three fits to the 
$Z$--$\log\Sigma_\star$ relation across 
all 3608 valid bins: a linear fit 
(RMSE $= 0.1014$\,dex), a quadratic fit 
(RMSE $= 0.1011$\,dex, improvement 
$0.3$ per cent), and a piecewise linear 
fit with a break at the median 
$\log\Sigma_\star$ 
(RMSE $= 0.1008$\,dex, improvement 
$0.6$ per cent). The non-linear fits are 
negligibly better than the linear fit, 
and we therefore retain the simpler 
linear parametrisation.

Second, we tested whether the residuals 
of the linear fit show a systematic trend 
with radius. We find a weak but 
significant positive trend of 
$+0.017$\,dex\,$R_e^{-1}$ in the median 
residuals across the ten radial bins 
($r = 0.939$, $p < 0.001$). This 
artificial trend arises because the 
linear fit slightly overestimates 
metallicity in the inner 
high-$\Sigma_\star$ bins and 
underestimates it in the outer 
low-$\Sigma_\star$ bins. We correct for 
this by subtracting 
$+0.017 \times (r/R_e)$ from each 
$\Delta Z$ value before fitting the 
residual gradient slopes. After this 
correction the median $\Delta Z$ slope 
changes from $+0.044$ to 
$+0.001$\,dex\,$R_e^{-1}$, consistent 
with zero, and the concentration 
correlation is completely unchanged 
($r = 0.304$, $p < 0.001$ before and 
after correction). All $\Delta Z$ slopes 
reported in the main text use this 
corrected definition.

\section{Metallicity and Stellar Mass 
Surface Density}
\label{app:zsigma}
\begin{figure}
\centering
\includegraphics[width=\hsize]{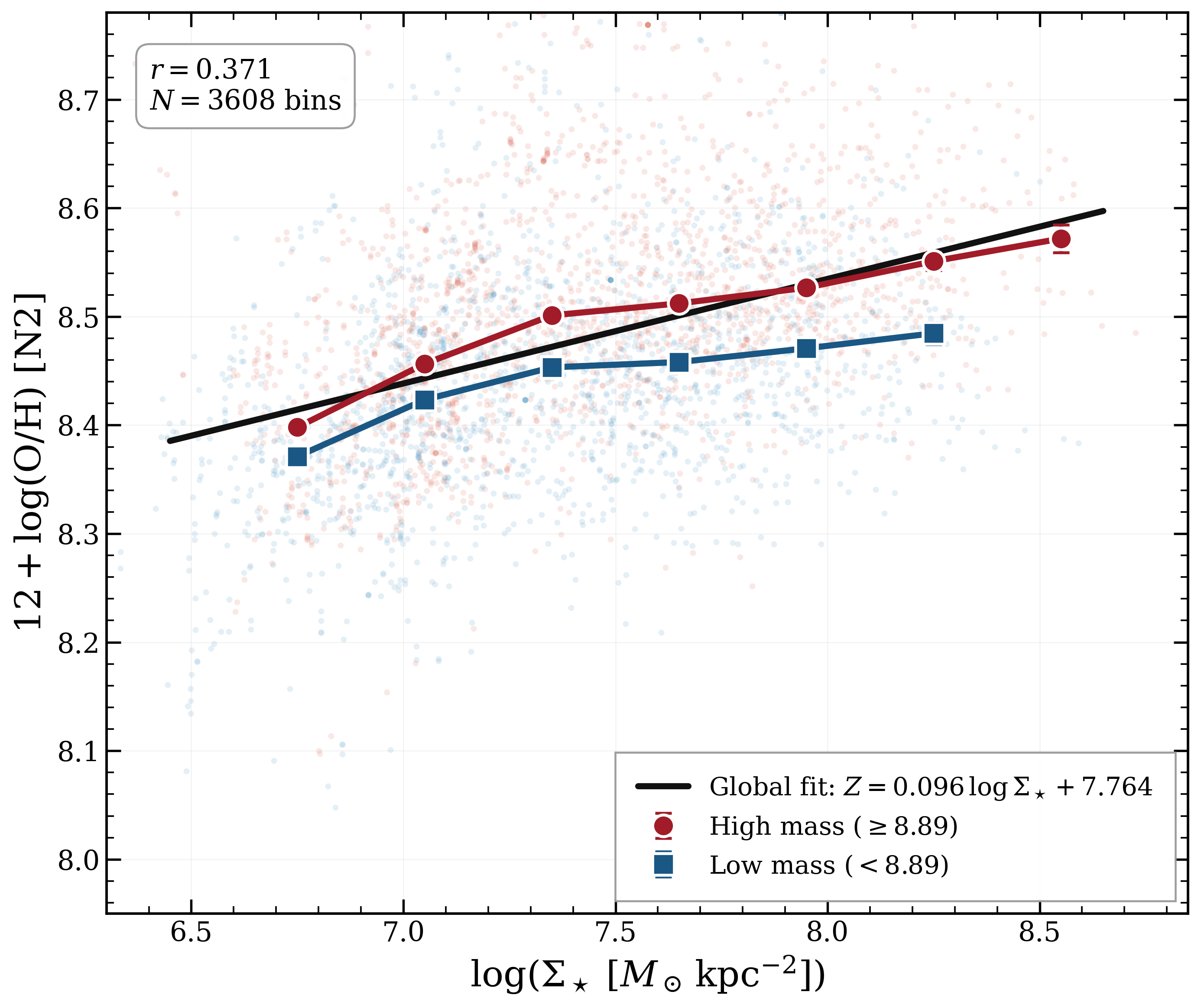}
\caption{Gas-phase metallicity as a function 
of local stellar mass surface density 
$\Sigma_{\star}$ for all 3608 valid radial 
bins from all 382 galaxies. Points are 
colour coded by stellar mass group: high 
mass ($\log(M_{\star}/M_{\odot}) \geq 8.89$, 
red, $N = 191$ galaxies) and low mass 
($\log(M_{\star}/M_{\odot}) < 8.89$, blue, 
$N = 191$ galaxies). The solid line shows 
the global linear fit 
$Z = 0.0963\,\log(\Sigma_{\star}) + 7.7643$ 
(Pearson $r = 0.371$, 
$p = 2.0 \times 10^{-118}$, $N = 3608$ 
bins), used to define the residual 
metallicity $\Delta Z$ in 
Section~\ref{sec:residual}. The filled 
circles and squares show the running median 
for the high and low mass groups 
respectively. The two mass groups are offset 
in metallicity at all values of 
$\log(\Sigma_{\star})$, demonstrating that 
total stellar mass influences chemical 
enrichment independently of local stellar 
mass surface density.}
\label{fig:zsigma}
\end{figure}

Figure~\ref{fig:zsigma} shows the 
relationship between gas-phase metallicity 
and local stellar mass surface density 
$\Sigma_{\star}$ for all 3608 valid radial 
bins from all 382 galaxies. The significant 
positive correlation ($r = 0.371$, 
$p = 2.0 \times 10^{-118}$) confirms that 
denser regions are on average more metal 
rich, consistent with more chemically 
evolved stellar populations at higher 
surface densities. The two mass groups are 
systematically offset in metallicity at 
fixed $\Sigma_{\star}$, motivating the 
residual metallicity analysis described in 
Section~\ref{sec:residual}. The linearity 
of this relation and the small systematic 
correction applied to the residuals are 
discussed in Appendix~\ref{app:nonlinear}.
To quantify whether the two mass groups 
are genuinely offset in $\Delta Z$ at 
fixed $\Sigma_\star$, we compared the 
$\Delta Z$ distributions for all valid 
bins in the high-mass 
($\log(M_\star/M_\odot) \geq 8.89$, 
1827 bins) and low-mass 
($\log(M_\star/M_\odot) < 8.89$, 
1781 bins) groups. The high-mass group 
has a median $\Delta Z = -0.004$ and 
the low-mass group a median of 
$-0.051$, an offset of $+0.047$\,dex. 
A Mann-Whitney test confirms the offset 
is highly significant ($p < 0.001$), 
demonstrating that total stellar mass 
influences chemical enrichment 
independently of local stellar mass 
surface density.

\section{Diffuse Ionized Gas Test}
\label{app:dig}

Diffuse ionized gas (DIG) can bias 
emission-line metallicity measurements 
because its line ratios differ from those 
of star-forming \HII\ regions. To assess 
whether DIG contamination introduces a 
systematic bias into our gradient 
measurements, we performed a half-ellipse 
test. For each galaxy we split the IFU 
spaxels into two halves along the minor 
axis and measured the metallicity gradient 
slope independently on each half. In the 
absence of systematic DIG contamination, 
the two halves should give consistent 
gradients with differences distributed 
symmetrically around zero.

\begin{figure}
\centering
\includegraphics[width=\hsize]{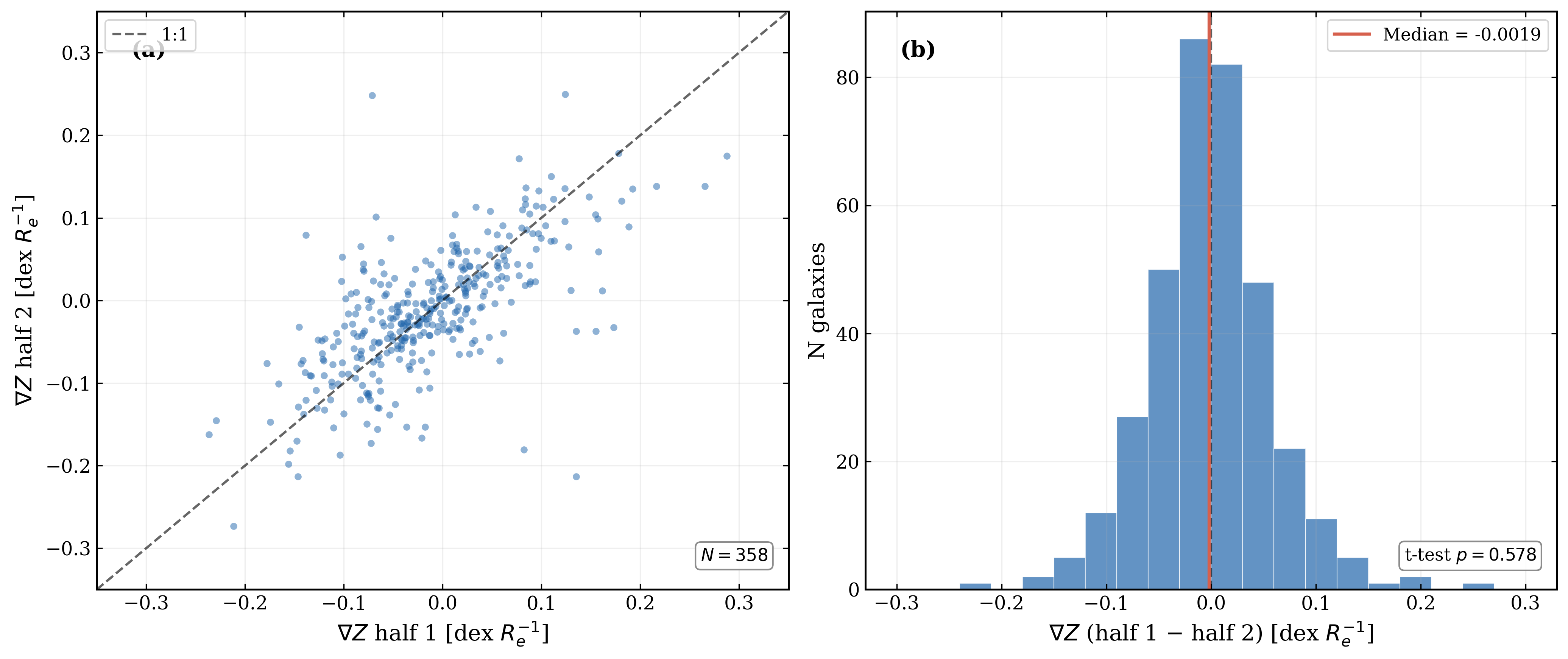}
\caption{Half-ellipse test for diffuse 
ionized gas contamination. For each galaxy 
the IFU is split into two halves along the 
minor axis and the metallicity gradient 
slope is measured independently on each 
half. \textit{Panel (a):} gradient slope 
measured on half 1 versus half 2 for 358 
galaxies. The dashed line shows the 1:1 
relation. The strong correlation 
($r = 0.631$, $p < 0.001$) confirms 
physically consistent measurements across 
both halves. \textit{Panel (b):} histogram 
of the difference between the two half 
slopes. The distribution is symmetric 
around zero (median $= -0.002$\,dex\, 
$R_e^{-1}$, $t$-test $p = 0.578$), 
confirming the absence of a systematic 
bias from DIG contamination.}
\label{fig:dig}
\end{figure}

We applied this test to all 382 galaxies, 
obtaining valid gradient measurements for 
both halves in 358 galaxies. The two halves 
show a strong positive correlation 
($r = 0.631$, $p < 0.001$), confirming that 
the gradient measurements are physically 
consistent across both sides of each galaxy. 
The differences between the two halves have 
a median of $-0.002$\,dex\,$R_e^{-1}$, 
consistent with zero (one-sample $t$-test 
$p = 0.578$), and 91 per cent of galaxies 
show absolute differences smaller than 
$0.10$\,dex\,$R_e^{-1}$. The results are 
shown in Figure~\ref{fig:dig}.

These results confirm that DIG contamination 
does not introduce a systematic bias into 
our gradient measurements.

\section{Monte Carlo Error Propagation}
\label{app:mc}

Our primary correlation result is between 
light concentration $C$ and residual 
gradient slope $\nabla\Delta Z$. To 
quantify how sensitive this result is to 
measurement uncertainties in the gradient 
slopes, we performed a Monte Carlo 
analysis on both the full sample and the 
stricter resolution subsample.

\begin{figure}
\centering
\includegraphics[width=\hsize]{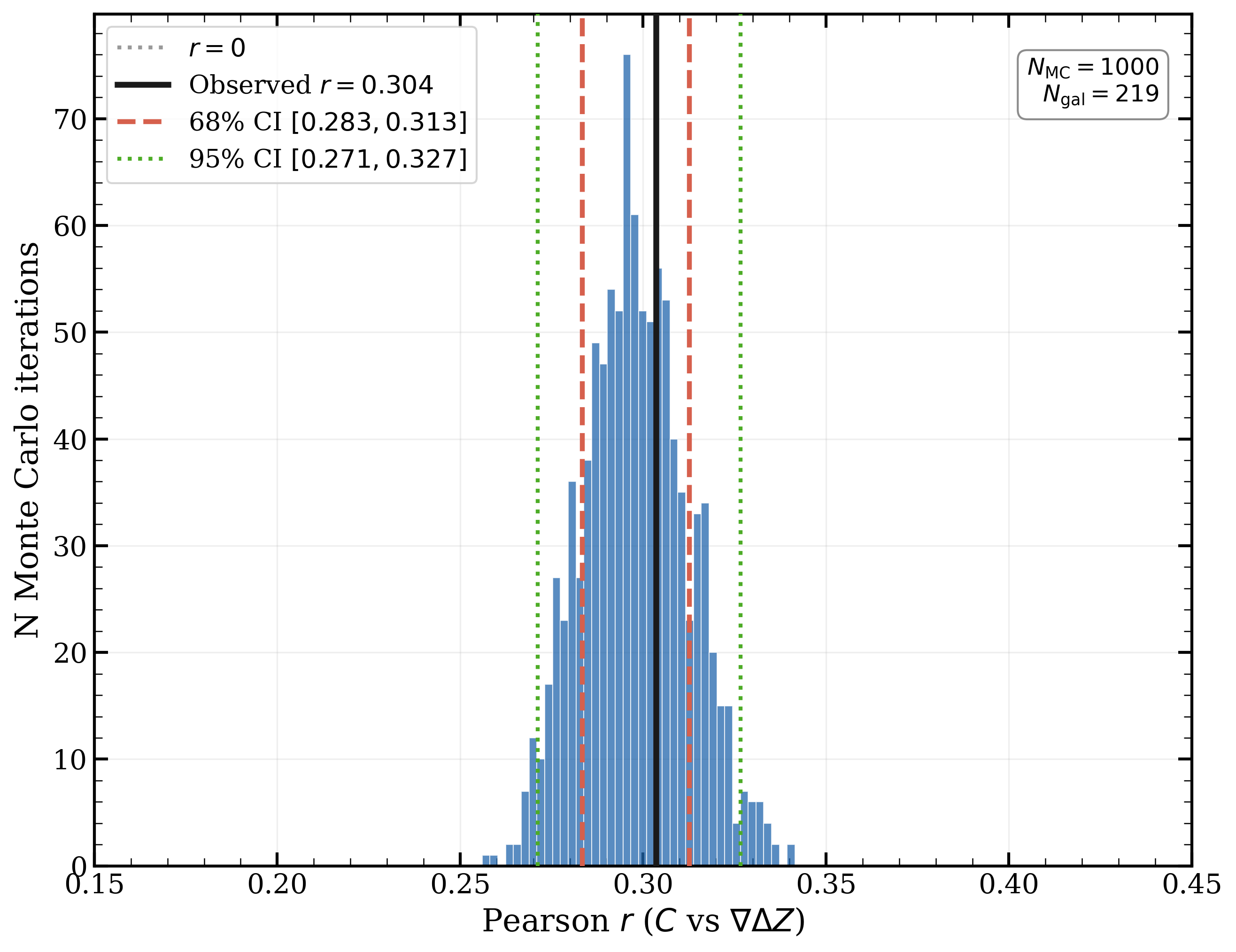}
\caption{Monte Carlo distribution of the 
Pearson correlation coefficient between 
light concentration $C$ and residual 
gradient slope $\nabla\Delta Z$ for the 
full sample ($N = 219$ galaxies with 
valid $C$ and $\nabla\Delta Z$ 
measurements), based on 1000 iterations 
in which each $\nabla\Delta Z$ value is 
perturbed within its measurement 
uncertainty $\sigma_{\nabla\Delta Z}$. 
The vertical solid black line marks the 
observed value ($r = 0.304$), the red 
dashed lines mark the 68 per cent 
confidence interval $[+0.283, +0.313]$, 
and the green dotted lines mark the 
95 per cent confidence interval 
$[+0.271, +0.327]$. The distribution is 
entirely positive and well separated 
from zero, confirming the robustness of 
the concentration correlation.}
\label{fig:mc}
\end{figure}

For the full sample of 366 clean galaxies, 
219 have both a valid concentration 
measurement and a well-constrained gradient 
slope. Of the 230 galaxies with a 
concentration measurement, 11 fail the 
gradient quality cut 
$\sigma_{\nabla\Delta Z} \leq 
0.05$\,dex\,$R_{\rm e}^{-1}$, leaving 219. For each 
of 1000 iterations we perturbed the 
$\nabla\Delta Z$ value of each galaxy by 
a random draw from a Gaussian distribution 
centred on the measured value with standard 
deviation equal to 
$\sigma_{\nabla\Delta Z}$, and recomputed 
the Pearson correlation coefficient with 
$C$. The resulting distribution is shown 
in Figure~\ref{fig:mc}.

For the full sample ($N = 219$, 
$r = 0.304$) the Monte Carlo mean is 
$r = 0.298$ with a 68 per cent confidence 
interval of $[+0.283, +0.313]$ and a 
95 per cent confidence interval of 
$[+0.271, +0.327]$. The entire 
distribution lies well above zero, 
confirming the robustness of the 
concentration correlation to measurement 
uncertainties.

For the stricter resolution subsample 
($R_{\rm e} \geq 2\,\theta_{\rm PSF}$), 
137 of the 254 galaxies have a valid 
concentration measurement and 
well-constrained gradient slope 
($r = 0.234$); the Monte Carlo mean is $r = 0.239$, equal to the 
observed value within rounding, with a 68 per cent 
confidence interval of 
$[+0.214, +0.256]$ and a 95 per cent 
confidence interval of 
$[+0.191, +0.276]$. Both distributions 
are entirely positive, confirming that 
the concentration correlation is robust 
to measurement uncertainties in both 
samples.

\section{N2 and O3N2 Metallicity 
Calibration Comparison}
\label{app:calibration}
We use the N2 strong-line calibration of 
\citet{PettiniPagel2004} as our primary 
metallicity indicator throughout this work. 
To validate this choice we compare the 
performance of N2 and O3N2 in recovering 
the mass-metallicity relation (MZR) for our 
sample.

\begin{figure}
\centering
\includegraphics[width=\hsize]{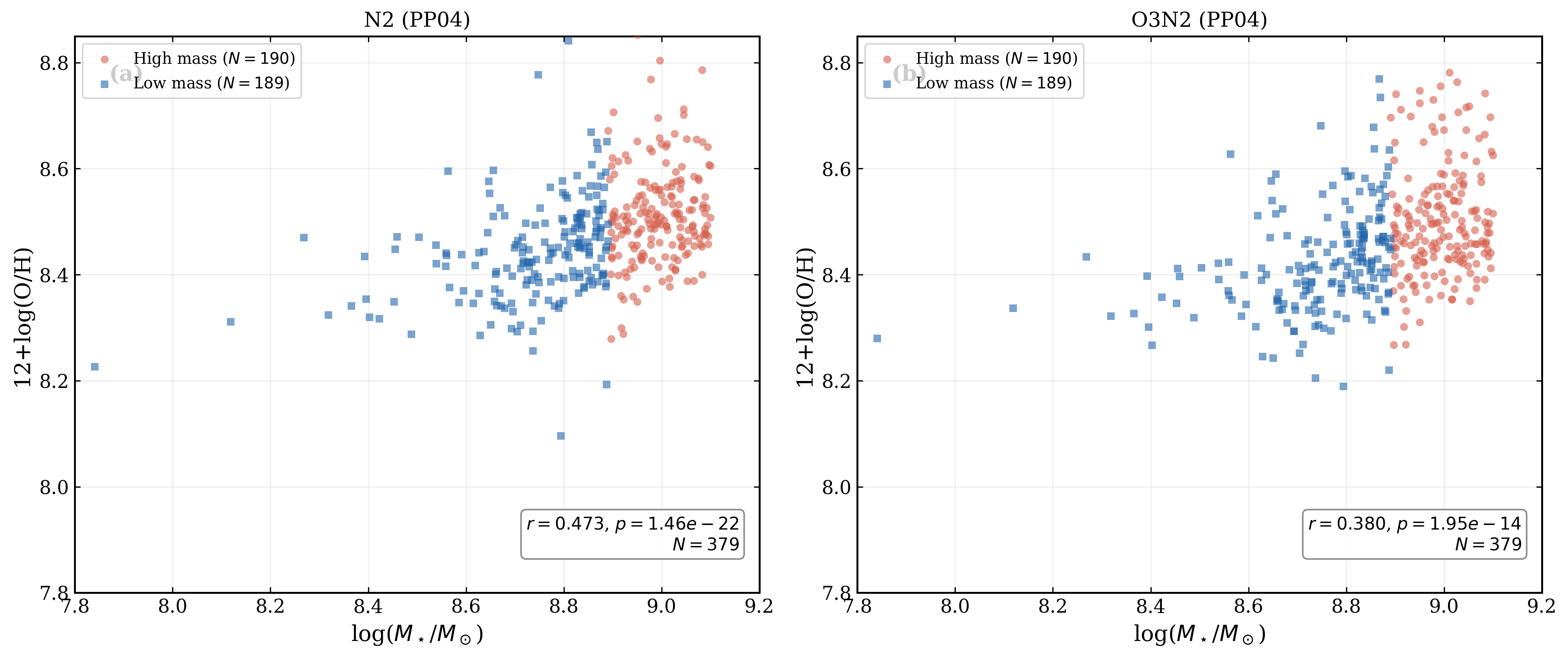}
\caption{Mass-metallicity relation for 
379 galaxies in our sample with valid 
metallicity measurements within 
$1\,R_e$, using the N2 
(\textit{left}, Pearson $r = 0.473$, 
$p < 0.001$) and O3N2 (\textit{right}, 
$r = 0.380$, $p < 0.001$) calibrations 
of \citet{PettiniPagel2004}. Points are 
colour coded by stellar mass group: high 
mass (red circles) and low mass (blue 
squares). Both calibrations recover a 
significant mass-metallicity relation, 
with N2 giving a stronger and cleaner 
correlation. The N2 calibration is 
therefore adopted as the primary 
metallicity indicator throughout this 
work.}
\label{fig:calib}
\end{figure}

Figure~\ref{fig:calib} shows the MZR for 
the 379 galaxies with valid metallicity 
measurements within $1\,R_e$ in both 
calibrations. One galaxy (8934-9102) is 
excluded because its N2-derived 
metallicity is unphysically high 
($12 + \log(\mathrm{O/H}) = 9.20$, 
exceeding the calibration range of 
\citealt{PettiniPagel2004}), most likely 
reflecting a failed emission-line fit 
rather than a true abundance; its O3N2 
value is also unphysical ($-1.42$). The 
N2 calibration recovers a clear and 
significant MZR (Pearson $r = 0.473$, 
$p < 0.001$, $N = 379$). The O3N2 
calibration also recovers a significant 
MZR but with a weaker correlation 
($r = 0.380$, $p < 0.001$, $N = 379$), 
consistent with the known larger scatter 
of O3N2 at low metallicities 
\citep{KewleyEllison2008}. The N2 
calibration is therefore the more reliable 
indicator for our low-mass sample and is 
adopted throughout the main analysis.

%\FloatBarrier

\onecolumn

\section{Master Table}
\label{app:table}
\begin{center}
\captionof{table}{Key properties of the 
382 low-mass galaxies in our final 
sample, sorted by stellar mass. Only the 
first 50 entries are shown here; the 
full table is available as online 
supplementary material. Columns give the 
MaNGA plate-IFU identifier, spectroscopic 
redshift, log stellar mass, effective 
radius, spatial resolution ratio 
$R_{\rm e}/\theta_{\rm PSF}$, metallicity 
gradient slope with $1\sigma$ uncertainty, 
corrected residual metallicity gradient 
slope with $1\sigma$ uncertainty, and 
visual classification. Three galaxies (11020-12701, 8554-12704, 8150-3701) have catalogue redshifts below the nominal selection range $0.01 \leq z \leq 0.05$; see Section ~\ref{sec:redshift}}
\label{tab:master}
\vspace{0.5em}
\setlength{\tabcolsep}{5pt}
\small
\begin{tabular}{lcccccrl}
\hline\hline
PlateIFU & $z$ & $\log M_{\star}$ & 
$R_{\rm e}$ & 
$R_{\rm e}/\theta_{\rm PSF}$ & 
$\nabla Z \pm \sigma$ & 
$\nabla\Delta Z \pm \sigma$ & 
Class \\
 & & ($M_{\odot}$) & (\arcsec) & & 
(dex\,$R_{\rm e}^{-1}$) & 
(dex\,$R_{\rm e}^{-1}$) & \\
\hline

11020-12701 & 0.0002 & 7.22 & 32.72 & 13.09 & $+0.077\pm0.018$ & $\dots$ & ISOLATED \\
8150-3701 & 0.0017 & 7.57 & 25.60 & 10.24 & $+0.164\pm0.002$ & $\dots$ & DISTURBED \\
8554-12704 & 0.0005 & 7.79 & 29.68 & 11.87 & $+0.074\pm0.043$ & $\dots$ & ISOLATED \\
9049-6101 & 0.0139 & 7.84 & 7.35 & 2.94 & $-0.094\pm0.023$ & $-0.052\pm0.027$ & ISOLATED \\
8981-12705 & 0.0057 & 8.12 & 6.10 & 2.44 & $-0.076\pm0.013$ & $-0.023\pm0.016$ & ISOLATED \\
8619-12705 & 0.0116 & 8.27 & 7.87 & 3.15 & $-0.085\pm0.016$ & $-0.043\pm0.013$ & DISTURBED \\
9493-12702 & 0.0122 & 8.32 & 7.14 & 2.86 & $-0.081\pm0.008$ & $-0.037\pm0.009$ & ISOLATED \\
8546-12702 & 0.0125 & 8.37 & 6.48 & 2.59 & $+0.019\pm0.012$ & $+0.072\pm0.014$ & ISOLATED \\
9502-6102 & 0.0173 & 8.39 & 7.59 & 3.03 & $+0.003\pm0.016$ & $+0.025\pm0.022$ & ISOLATED \\
8565-6104 & 0.0094 & 8.40 & 4.72 & 1.89 & $+0.087\pm0.012$ & $+0.138\pm0.012$ & ISOLATED \\
9093-1902 & 0.0220 & 8.40 & 3.04 & 1.22 & $+0.017\pm0.012$ & $+0.053\pm0.015$ & ISOLATED \\
8993-6101 & 0.0196 & 8.42 & 7.48 & 2.99 & $-0.037\pm0.008$ & $-0.047\pm0.008$ & ISOLATED \\
8619-12704 & 0.0206 & 8.45 & 5.57 & 2.23 & $-0.032\pm0.008$ & $+0.005\pm0.013$ & ISOLATED \\
8719-9101 & 0.0178 & 8.46 & 12.38 & 4.95 & $+0.028\pm0.050$ & $+0.084\pm0.047$ & DISTURBED \\
8082-3701 & 0.0213 & 8.46 & 2.87 & 1.15 & $+0.036\pm0.007$ & $+0.078\pm0.008$ & DISTURBED \\
8982-12702 & 0.0250 & 8.49 & 6.81 & 2.72 & $-0.095\pm0.011$ & $-0.066\pm0.006$ & ISOLATED \\
8552-3701 & 0.0183 & 8.50 & 4.46 & 1.78 & $+0.104\pm0.015$ & $+0.154\pm0.018$ & DISTURBED \\
9886-6102 & 0.0219 & 8.54 & 6.62 & 2.65 & $+0.017\pm0.016$ & $+0.065\pm0.017$ & ISOLATED \\
8338-1901 & 0.0215 & 8.54 & 3.19 & 1.28 & $+0.106\pm0.017$ & $+0.157\pm0.019$ & DISTURBED \\
8657-6104 & 0.0175 & 8.56 & 5.17 & 2.07 & $-0.027\pm0.010$ & $+0.016\pm0.009$ & DISTURBED \\
8149-6103 & 0.0221 & 8.56 & 5.36 & 2.14 & $-0.072\pm0.015$ & $-0.048\pm0.010$ & DISTURBED \\
11022-12702 & 0.0228 & 8.56 & 6.00 & 2.40 & $+0.089\pm0.017$ & $+0.124\pm0.014$ & ISOLATED \\
9494-6102 & 0.0157 & 8.56 & 6.04 & 2.41 & $+0.173\pm0.033$ & $+0.199\pm0.038$ & ISOLATED \\
8996-6104 & 0.0214 & 8.57 & 6.62 & 2.65 & $-0.011\pm0.008$ & $-0.020\pm0.008$ & ISOLATED \\
9001-12704 & 0.0212 & 8.59 & 8.94 & 3.58 & $-0.030\pm0.015$ & $+0.010\pm0.020$ & ISOLATED \\
8447-1901 & 0.0202 & 8.59 & 3.16 & 1.27 & $-0.026\pm0.005$ & $+0.020\pm0.004$ & ISOLATED \\
8459-9102 & 0.0170 & 8.59 & 7.75 & 3.10 & $+0.021\pm0.016$ & $+0.063\pm0.017$ & ISOLATED \\
8553-12703 & 0.0136 & 8.61 & 12.40 & 4.96 & $+0.198\pm0.053$ & $+0.241\pm0.056$ & DISTURBED \\
9886-3701 & 0.0216 & 8.62 & 12.84 & 5.13 & $-0.151\pm0.091$ & $-0.063\pm0.098$ & DISTURBED \\
8563-3704 & 0.0192 & 8.63 & 3.79 & 1.52 & $+0.047\pm0.007$ & $+0.095\pm0.007$ & ISOLATED \\
8987-6101 & 0.0219 & 8.63 & 5.10 & 2.04 & $-0.069\pm0.008$ & $-0.048\pm0.006$ & ISOLATED \\
8241-6101 & 0.0209 & 8.63 & 5.82 & 2.33 & $+0.104\pm0.038$ & $+0.162\pm0.040$ & DISTURBED \\
8566-6103 & 0.0103 & 8.64 & 4.69 & 1.88 & $-0.018\pm0.005$ & $+0.020\pm0.009$ & ISOLATED \\
8564-1901 & 0.0196 & 8.64 & 3.35 & 1.34 & $+0.044\pm0.008$ & $+0.079\pm0.011$ & DISTURBED \\
8439-1901 & 0.0164 & 8.65 & 3.21 & 1.28 & $+0.097\pm0.007$ & $+0.143\pm0.006$ & ISOLATED \\
8135-6101 & 0.0109 & 8.65 & 6.84 & 2.74 & $+0.077\pm0.016$ & $+0.112\pm0.014$ & DISTURBED \\
8936-6104 & 0.0142 & 8.65 & 6.16 & 2.46 & $+0.094\pm0.017$ & $+0.158\pm0.020$ & DISTURBED \\
10517-12704 & 0.0132 & 8.66 & 9.39 & 3.76 & $-0.289\pm0.132$ & $-0.128\pm0.107$ & DISTURBED \\
8991-3703 & 0.0191 & 8.66 & 4.41 & 1.77 & $-0.006\pm0.004$ & $+0.041\pm0.004$ & ISOLATED \\
10496-12703 & 0.0189 & 8.66 & 12.05 & 4.82 & $-0.068\pm0.009$ & $-0.045\pm0.008$ & DISTURBED \\
11977-12704 & 0.0211 & 8.66 & 10.06 & 4.02 & $-0.033\pm0.015$ & $+0.025\pm0.022$ & DISTURBED \\
9001-9102 & 0.0214 & 8.66 & 2.64 & 1.05 & $+0.073\pm0.003$ & $+0.106\pm0.005$ & ISOLATED \\
8567-6102 & 0.0227 & 8.66 & 3.81 & 1.52 & $+0.013\pm0.008$ & $+0.041\pm0.011$ & DISTURBED \\
8552-6101 & 0.0180 & 8.66 & 6.54 & 2.62 & $-0.050\pm0.010$ & $-0.042\pm0.005$ & DISTURBED \\
9485-9102 & 0.0188 & 8.67 & 7.91 & 3.16 & $-0.002\pm0.011$ & $-0.011\pm0.007$ & DISTURBED \\
8716-12701 & 0.0190 & 8.67 & 7.45 & 2.98 & $-0.009\pm0.001$ & $-0.015\pm0.002$ & DISTURBED \\
8987-6103 & 0.0217 & 8.67 & 7.52 & 3.01 & $-0.063\pm0.007$ & $-0.025\pm0.011$ & ISOLATED \\
9500-3702 & 0.0216 & 8.68 & 4.60 & 1.84 & $+0.069\pm0.010$ & $+0.121\pm0.015$ & DISTURBED \\
8323-12702 & 0.0230 & 8.68 & 8.99 & 3.60 & $+0.207\pm0.052$ & $+0.250\pm0.054$ & ISOLATED \\
8551-6101 & 0.0188 & 8.68 & 7.30 & 2.92 & $-0.052\pm0.008$ & $-0.035\pm0.005$ & DISTURBED \\
\hline
\end{tabular}
\end{center}
\end{appendix}
\end{document}